\documentclass[a4paper,11pt]{article}
\pdfoutput=1
\usepackage{jcappub}
\usepackage{txfonts}
\usepackage{threeparttable}
\usepackage{aas_macros}

\newcommand{\ud}{\mathrm{d}}
\renewcommand\({\left(}
\renewcommand\){\right)}
\renewcommand\[{\left[}
\renewcommand\]{\right]}

\def\Or{\Omega_{\rm{r}}}
\def\Om{\Omega_{\rm{m}}}
\def\Ok{\Omega_{\rm{k}}}
\def\Ol{\Omega_{\Lambda}}
\def\Obh{\Omega_{\rm{b}}h^2}
\def\Ob{\Omega_{\rm{b}}}
\def\Onu{\Omega_\nu}

\def\Och{\Omega_{\rm{DM}}h^2}
\def\Oc{\Omega_{\rm{DM}}}
\def\ns{n_{\rm s}}
\def\As{A_{\rm s}}

\def\thA{\theta_{\rm A}}

\newcommand{\neff}{\ensuremath{N_\mathrm{eff}}}
\newcommand{\mnu}{\ensuremath{\sum{m_{\nu}}}}

\newcommand{\Map}[1]{\left<M^2_\mathrm{ap}\right>( #1 )}
\newcommand{\map}{\ensuremath{\left<M^2_\mathrm{ap}\right>}}
\newcommand{\chiH}{\ensuremath{\chi_\mathrm{H}}}

\title{Observational constraints on cosmic neutrinos and dark energy revisited}

\author[a,b]{Xin Wang,}
\author[a,c]{Xiao-Lei Meng,}
\author[a,d]{Tong-Jie Zhang,}
\author[e,f]{HuanYuan Shan,}
\author[g]{Yan Gong,}
\author[e,f]{Charling Tao,}
\author[c]{Xuelei Chen,}
\author[b]{Y. F. Huang}

\affiliation[a]{Department of Astronomy, Beijing Normal University, Beijing 100875, China}
\affiliation[b]{School of Astronomy and Space Science, Nanjing University, Nanjing, 210093, China}
\affiliation[c]{National Astronomical Observatories, Chinese Academy of Sciences, Beijing 100012, China}
\affiliation[d]{Kavli Institute for Theoretical Physics China, CAS, Beijing 100190, China}
\affiliation[e]{Department of Physics and Tsinghua Center for Astrophysics, Tsinghua University, Beijing, 100084, China}
\affiliation[f]{Centre de Physique des Particules de Marseille, CNRS/IN2P3-Luminy and Universit\'e de la M\'editerran\'ee, Case 907, F-13288 Marseille Cedex 9, France}
\affiliation[g]{Department of Physics and Astronomy, University of California, Irvine, CA 92697}

\emailAdd{tjzhang@bnu.edu.cn}

\abstract{ 
Using several cosmological observations, i.e. the cosmic microwave background anisotropies (WMAP), the weak gravitational lensing (CFHTLS), the measurements of baryon acoustic oscillations (SDSS+WiggleZ), the most recent observational Hubble parameter data, the Union2.1 compilation of type Ia supernovae, and the HST prior, we impose constraints on the sum of neutrino masses ($\mnu$), the effective number of neutrino species ($\neff$) and dark energy equation of state ($w$), individually and collectively. We find that a tight upper limit on $\mnu$ can be extracted from the full data combination, if $\neff$ and $w$ are fixed. However this upper bound is severely weakened if $\neff$ and $w$ are allowed to vary. This result naturally raises questions on the robustness of previous strict upper bounds on $\mnu$, ever reported in the literature. The best-fit values from our most generalized constraint read $\mnu=0.556^{+0.231}_{-0.288}\rm~eV$, $\neff=3.839\pm0.452$, and $w=-1.058\pm0.088$ at 68\% confidence level, which shows a firm lower limit on total neutrino mass, favors an extra light degree of freedom, and supports the cosmological constant model. The current weak lensing data are already helpful in constraining cosmological model parameters for fixed $w$. The dataset of Hubble parameter gains numerous advantages over supernovae when $w=-1$, particularly its illuminating power in constraining $\neff$. As long as $w$ is included as a free parameter, it is still the standardizable candles of type Ia supernovae that play the most dominant role in the parameter constraints.
}


\begin{document}
\maketitle
\flushbottom

\section{Introduction}\label{sec:intro}
One of the most intriguing mysteries in modern cosmology
is the property of cosmic neutrinos. Directly from the oscillation 
experiments \citep{2004NJPh....6..122M}, 
it is known that
\begin{eqnarray}	\label{eq:nu_spli_mass}
\Delta m^2_{\rm atm} = |\Delta m^2_{31}|=(2.2^{+0.7}_{-0.5})\times 10^{-3} \rm eV^2\nonumber\\
\Delta m^2_{\rm sun} = \Delta m^2_{21}=(8.1^{+0.6}_{-0.6})\times 10^{-5} \rm eV^2,
\end{eqnarray}
where $\Delta m^2_{ij}=m_i^2-m_j^2$ is the squared mass difference 
between two neutrino eigenstates. This result naturally leads to three scenarios 
for the mass splitting of the standard three flavor neutrinos \citep{2006ARNPS..56..137H}:

\hspace{0.2cm}(I) Normal Hierarchy --- $m_1\sim0$, $m_2\sim\Delta m _{\rm sun}$, and $m_3\sim\Delta m_{\rm atm}$;

\hspace{0.2cm}(II) Inverted Hierarchy --- $m_1\sim m_2\sim\Delta m _{\rm atm}$, and $m_3\sim0$;

\hspace{0.2cm}(III) Total Degeneracy  --- $m_1\sim m_2\sim m_3\gg \Delta m_{\rm atm}$.\\
Although the absolute values of neutrino masses are beyond the reach of oscillation experiments, 
other ground-based experiments, e.g. measurements of 
tritium beta decay \citep{2005EPJC...40..447K} and neutrinoless double beta 
decay \citep{2001EPJA...12..147K}, provide some hints on the overall mass scale.
Nevertheless, the most compelling bounds on the total mass of neutrinos ($\mnu$) come from the cosmos (for recent reviews, see
Refs.~\citep{2010PrPNP..65..185H, 2011APh....35..177A}).
Firstly, the anisotropies of the cosmic microwave background (CMB) are sensitive to neutrino masses \citep{2006PhR...429..307L}.
As massive neutrinos become non-relativistic, 
their contributions to the energy constituent changes from radiation-like to matter-like.
Provided that these neutrinos are massive enough, they will slow down to 
non-relativistic speed even before the recombination epoch. This slight modification in expansion history affects acoustic peaks.
Small scale anisotropies are meanwhile boosted, due to the enhancement of photon energy 
density fluctuations by the diminishing gravitational potentials \citep{1996ApJ...467...10D}. 
The latest observations of WMAP seven-year pin down the 95\% upper limit of $\mnu$ 
to $1.3\rm~eV$ \citep{2011ApJS..192...18K}, 
while Planck and other next-term probes are presumably capable 
of pushing forward the sensitivity to $\mnu\sim 0.2-0.03\rm~eV$, 
\citep{2006PhRvD..73d5021L, 2009AIPC.1141...10B, 2012JCAP...03..023C}.
Secondly, massive neutrinos affect the evolution of matter perturbation in a particular way:
they erase the density contrast at wavelengths smaller than their characteristic 
free-streaming scale, resulting in a suppression of linear matter 
power spectrum as given by \citep{1998PhRvL..80.5255H},
\begin{equation} \label{eq:deltaP}
\frac{\Delta P_{\rm lin}(k)}{P_{\rm lin}(k)}\sim-8\frac{\Onu}{\Om}.
\end{equation}
with $\Onu=\mnu/\(93.8~h^2~\rm eV\)$ being the neutrino matter density fraction \citep{2006ARNPS..56..137H}.
Thus the constraint on the sum of neutrino masses greatly benefits from accurate 
measurements of matter power spectrum. A great deal of effort has been dedicated to reducing
the upper limit of $\mnu$ from a number of large scale structure projects,
e.g. 2dFGRS, SDSS, WiggleZ, CFHTLS galaxy surveys \citep{2002PhRvL..89f1301E, 2003JCAP...05..004H,
2003MNRAS.346..593A, 2004JCAP...04..008H, 2004PhLB..595...55B, 2004PhRvD..69l3007C, 2004PhRvD..69j3501T,
2005PhRvD..71j3515S, 2006JCAP...10..014S, 2006JCAP...06..019G, 2007JCAP...07..010Z, 2008PhRvD..78c3010F, 
2008PhRvD..78h3524X, 2010JCAP...03..015S, 2010MNRAS.404...60R, 2010PhRvL.105c1301T, 
2011PhRvD..83d3529S, 2012PhRvD..85h1101R, 2012JCAP...06..010X}.
The present record gives $\mnu < 0.26\rm~eV$ at 95\% confidence level (CL),
using the combination of WMAP seven-year data, SDSS DR8 LRG angular spectra together with the 
HST prior \citep{2012arXiv1201.1909D}. 

Another novel and independent tracer of matter clustering is weak gravitational lensing (WL), i.e.
the bending of light through intervening inhomogeneous mass distribution, which as a result 
generates small distortion (at the 1\% level) on the images of distant sources. 
Since the first detection of cosmic shear \citep{2000Natur.405..143W, 2000MNRAS.318..625B,
2000A&A...358...30V, 2000astro.ph..3338K}, WL has been repeatedly shown to be a 
powerful and precise approach to measure mass fluctuations and constrain cosmological models 
(see Ref.~\citep{2001PhR...340..291B} for a thorough review). 
The greatest advantage of WL is that it only relies on the total matter content 
along the line of sight, and is therefore free from the problematic modeling 
of galaxy-to-matter bias \citep{2010GReGr..42.2177H}. Using Fisher formalism, 
several studies have forecasted the error budget of future WL surveys constraining 
neutrino masses \citep{1999A&A...348...31C, 2003PhRvL..91d1301A, 2006JCAP...06..025H, 
2012PhRvD..86b3526J}. However, there are few groups who have put actual 
constraints on $\mnu$ from current observational results \citep{2009PhLB..675..164L}.
In particular, using up-to-date WL data of CFHTLS-T0003 and other cosmological probes,
Ref.~\citep{2009A&A...500..657T} and Ref.~\citep{2009PhRvD..79b3520I}
obtained the mutually consistent result, an 95\% upper bound of $\mnu<0.54\rm~eV$.

Known as cosmic chronometers, the observational Hubble parameter data (OHD) have 
recently gained increasing attention in their potential of measuring the geometry
and matter content of the universe \citep{2007MPLA...22...41Y, 2010AdAst2010E..81Z}. 
The Hubble expansion rates at different redshifts can be obtained via the differential-age technique 
\citep{2002ApJ...573...37J, 2010JCAP...02..008S}. Compared with the type Ia supernovae (SNIa) data, 
OHD have the virtue of being directly linked to the expansion history,
whereas what SNIa measure is luminosity distance, related to $H(z)$ through integration.
Using the latest OHD data, Ref.~\citep{2012JCAP...07..053M} has derived competitive upper
limits on the total neutrino mass, $\mnu<0.24\rm~eV$ at 68\% CL. 

Although there has been a vast scope of work on the constraints on $\mnu$ previously, 
most of these analyses do not seek appropriate treatments of correlations 
associated with $\mnu$ and other cosmological parameters. As argued by Ref.~\citep{2005PhRvL..95v1301H}, 
the degeneracy between $\mnu$ and the dark energy equation of state (EOS) parameter $w$ 
fatally weakens the constraints on $\mnu$. 
Except the role of dark energy, the effective number of neutrino species, i.e. $\neff$, 
remarkably lifts the bounds on $\mnu$, attributed to their correlation 
\citep{2006JCAP...11..016H, 2007JCAP...08..021H, 2010PhRvL.105r1301H}.
Standard primordial nucleosynthesis with weak interaction rate corrections yields $\neff=3.046$ \cite{2005NuPhB.729..221M}. 
Here this non-integer part originates from the non-thermal feature in neutrino distribution funtion due to partial heating during 
$e^{\pm}$ annihilations, including both finite temperature QED corrections and flavor oscillation effects 
\citep{1982PhRvD..26.2694D, 1994PhRvD..49..611H, 2002NuPhB.632..363D}.
However, as proven more and more firmly by the series of WMAP papers, $\neff$ seems to be much larger than the standard value
\citep{2007ApJS..170..377S, 2009ApJS..180..306D, 2011ApJS..192...18K}. 
Similar conclusions are also reached by other small scale CMB observations,
e.g. $\neff=4.6\pm0.8$ for ACT, and $\neff=3.86\pm0.42$ for SPT \citep{2011ApJ...739...52D, 2011ApJ...743...28K}.
The presence of extra light degrees of freedom, called ``dark radiation'', might be reasonably 
explained by some light sterile neutrinos, axions or majorons in thermal equilibrium after 
the QCD phase transition 
\citep{2007JCAP...05..008I, 2011PhRvD..84l3008A, 2012arXiv1206.0109A}.
As a consequence, most of the upper limits on $\mnu$, 
which are derived fixing $\neff=3.046$, seem too optimistic. We note that recently Ref.~\citep{2012arXiv1202.0005J} uses
several updated cosmological probes to deliver one of the most comprehensive 
investigations on the ($\mnu,~\neff,~w$) constraints and 
their correlations. However Ref.~\citep{2012arXiv1202.0005J} does not focus on comparing different sets of cosmological data,
and furthermore it does not use one of the most promising techniques, i.e. WL.
In this work, we would like to concentrate on the correlations between some key parameters including ($\mnu,~\neff,~w$), and also investigate how 
their individual values and degeneracies affect the constraints as well as dark matter clustering.
Particularly, we will study the role of WL, and compare the constraining ability of SNIa and OHD. 

The outline of our paper is as follows. Sect.~\ref{sec:data} introduces the
observational data in our analysis. Then we present the main constraint
results in Sect.~\ref{sec:constrain}. In Sect.~\ref{sec:discus} we investigate in detail
the impacts of the three key parameters ($\mnu,~\neff,~w$) on structure formation.
Finally Sect.~\ref{sec:conclu} presents our conclusion and discussion.

\section{Observables and Data}\label{sec:data}

In this section, we briefly introduce the cosmological probes selected for neutrino constraints. 
In the context of Friedmann-Robertson-Walker metric, it is simple to derive
the coordinate distance $r$ in terms of the radial distance $\chi$ as \citep{2002PhRvD..65f3001H}
\begin{equation} \label{eq:rchi}
r(\chi)= \left\{
\begin{array}{lc}
 \frac{c}{H_0\sqrt{\Ok}} \sinh\(\frac{H_0\sqrt{\Ok}}{c}\chi\) & \rm{if}~~\Ok >0 , \\
 \chi & \rm{if}~~\Ok=0, \\
 \frac{c}{H_0\sqrt{|\Ok|}} \sin\(\frac{H_0\sqrt{|\Ok|}}{c}\chi\) & \rm{if}~~\Ok <0 , \\
 \end{array} \right.
\end{equation}
with $\chi$ and $r$ also known as the comoving distance and the comoving angular 
diameter distance respectively \citep{2005ASPC..339..215H}. Here the comoving distance
at a certain redshift $z'$ is given by 
\begin{equation}
\chi(z')=\frac{c}{H_0}\int_0^{z'} \frac{\ud z}{E(z)},
\end{equation} 
where $E(z)$ denotes the expansion rate of the universe \citep{2009ApJS..180..330K}, i.e.
\begin{equation}\label{eq:expan}
E(z) = \frac{H(z)}{H_0} = \[\frac{\Or}{a^4}+\frac{\Om}{a^3}+\frac{\Ok}{a^2}+\frac{\Ol}{a^{3(1+w)}}\]^{1/2},
\end{equation}
with $a=1/(1+z)$ standing for the scale factor.
Into the above formulae, the property of cosmic neutrinos enters via
\begin{eqnarray}
\Om & = & \Oc + \Ob + \Onu, \\
\Or & = & \Omega_{\gamma}\[1+\frac{7}{8}\(\frac{4}{11}\)^{4/3}\neff\].
\end{eqnarray}
Besides $\Onu$ and $\neff$, here $\Om,~\Oc,~\Ob,~\Or,~\Omega_{\gamma},~\Ok,~\Ol$ corresponds to present-day density fractions of 
all the matter, cold dark matter, baryon, radiation energy, photon energy, curvature, and dark energy, respectively.
$H_0$ is the Hubble constant and $c$ serves as the speed of light.

Cosmic massive neutrinos directly influence the expansion history of the Universe. 
They also leave their signatures on structure formation and mass distribution, 
which can be investigated through large scale structure and WL data. 
To use these data, one needs to know the exact
theoretical prediction of the matter power spectrum, i.e. $P(k)$, defined as the Fourier transform 
of the two-point correlation function of the matter density contrast $\delta(r)=1-\bar{n}(r)/n(r)$.  
The linear perturbation theory provides reliable results of $P(k)$ at large scales, 
usually referred to as the linear regime \citep{1986ApJ...304...15B, 1995ApJ...455....7M, 1998ApJ...496..605E}.
However, as the wavenumber $k$ rises up toward $\sim 0.1-0.6 \rm~h/Mpc$, the linear perturbation theory starts 
to break down, and various effects of non-linear clustering take place \citep{2011MNRAS.410.1647A}.
The scale-invariant suppression on $P(k)$ caused by massive neutrinos (Eq.~\ref{eq:deltaP}) thus no longer works.
Fortunately, people have developed numerous ways to explore those small scale phenomena, e.g. the
one-loop perturbation theorem \citep{2008JCAP...10..035W, 2008PhRvL.100s1301S, 2009PhRvD..80h3528S}, 
which usually takes up the third order expansion of $P(k)$ as a close approximation,
the time renormalization group flow method \citep{2009JCAP...06..017L}, and the most widely quoted 
HALOFIT analytical formulae as the mapping algorithm calibrated by N-body simulations \citep{2003MNRAS.341.1311S},
\begin{equation}
\Delta^2(k)=\frac{k^3P(k)}{2\pi}=\Delta^2_{\rm{Q}}+\Delta^2_{\rm{H}},
\end{equation}
where the total power spectrum $\Delta^2(k)$ is split into a quasi-linear term
$\Delta^2_{\rm{Q}}$ and a pure non-linear term $\Delta^2_{\rm{H}}$.
However, three problematic issues are found to affect the neutrino damping effect 
reproduced by this method \citep{2012MNRAS.420.2551B}:

\hspace{0.2cm}(I) HALOFIT overpredicts the suppression in non-linear regime, with maximum
discrepancy occuring at $k\sim1~\rm h/Mpc$;

\hspace{0.2cm}(II) HALOFIT cannot capture the redshift dependence of maximum
suppresion's location in power spectrum;

\hspace{0.2cm}(III) at very small scales ($k>2\rm~h/Mpc$), HALOFIT underpredicts the non-linear power.\\

Ref.~\citep{2012MNRAS.420.2551B} conducted a more extensive suite of N-body simulations 
particularly designed for quantifying precisely the suppression from massive neutrinos,
and improved the original method into HALOFIT--$\nu$ algorithm,
\begin{equation}
{\Delta'}^2(k)=\({\Delta'}^2_{\rm{Q}}+{\Delta'}^2_{\rm{H}}\)\cdot (1+Q_{\nu}),
\end{equation}
with both quasi- and non-linear terms modified, and $Q_{\nu}$ accounts for additional non-linear growth of
the neutrino component \citep[cf. Ref.][appendix A]{2012MNRAS.420.2551B}.
For the sake of the accuracy of our constraints, we adopt HALOFIT--$\nu$ as 
the correction to the original $P_{\rm lin}(k)$ in presence of massive neutrinos,
calculated numerically through the hierarchical Boltzmann equations.

\subsection{Weak Gravitational Lensing}

After the calculations of matter power spectra, under Limber's approximation, 
the convergence power spectrum of WL can be computed from the integration of 
matter power spectra along line of sight (see e.g. Ref.~\citep{2008PhR...462...67M}),
\begin{equation}\label{eq:Pkappa}
P^{\kappa}_l = \frac{9H_0^4\Om^2}{4c^4}\int^{\chiH}_0\frac{\ud\chi}{a^2(\chi)}
\[\int^{\chiH}_{\chi}\ud\chi'n(\chi')\frac{r(\chi'-\chi)}{r(\chi')}\]^2
P\(\frac{l}{r(\chi)},\chi\),
\end{equation}
where $l=k\cdot r(\chi)$ represents the multipole moment and $n(\chi(z))$ accounts for the source redshift distribution.
In principle, $\chiH$ should cover the entire Hubble radius, yet we can still safely replace it by $\chi(z_{\rm lim})$, i.e.  
the comoving distance out to the survey limited redshift ($z_{\rm lim}$), 
where the source normalized number count ($n(z)$) is sufficiently decaying. Since massive neutrinos damp 
matter power spectrum at small scales, the convergence power of WL is also reduced.

For the actual survey, we choose the published CFHTLS-T0003 observations 
which have measured about $2\times10^6$ galaxies with $i_{\rm AB}$ magnitudes between 21.5 and 24.5, imaged on
an area of 57 square degree (35 square degree effectively) \citep{2008A&A...479....9F}. 
CFHTLS-T0003 has provided the community with the observational results of the shear correlation function $\xi_{\rm E,B}$, 
the shear top-hat variance $\left<|\gamma|^2\right>_{\rm E,B}$, and the aperture-mass variance $\map$.
Amongst these three types of WL measurements, $\map$ has the virtue of less systematics prone due to 
its unambiguous local decomposition \citep{2009A&A...497..677K}. Theoretically, the aperture-mass variance
is related to the convergence power spectrum via \citep{1998MNRAS.296..873S,2002A&A...396....1S},
\begin{equation}\label{eq:Map}
\Map{\theta} = \frac{288}{\pi\theta^4}\int_0^{\infty}\frac{\ud l}{l^3}
J^2_4\(l\theta\)P^{\kappa}_l,
\end{equation}
with $J_{\alpha}(x)$ being the Bessel function of the first kind. Due to the choice of WL data, 
we use the following parameterization for the source distribution, as recommended by Ref.~\citep{2008A&A...479....9F},
\begin{eqnarray}
n(z) \propto \frac{z^a+z^{ab}}{z^b+c},\\
\int^{z_{\rm lim}}_0 n(z)\ud z =1, \nonumber 
\end{eqnarray}
where $a,~b,~\rm{and}~c$ are nuisance parameters
\footnote{We directly take the best-fit values of $a,~b,~\rm{and}~c$ presented in Table 2 of Ref.~\citep{2008A&A...479....9F}.
We have numerically checked that the correlation coefficients between these nuisance parameters and cosmological parameters
are trivial enough to be neglected. The similar approach is adopted in Ref.~\citep{2009PhLB..675..164L}.}
and the survey limited redshift is assigned the value of $z_{\rm lim}=7$ (M. Kilbinger, private communication).

\subsection{Type Ia Supernovae}
The use of SNIa as standardizable candles provides a powerful probe to explore the properties of dark energy.
SNIa observations furnish one of the metric distances, i.e. the luminosity distance, 
\begin{equation}
D_{\rm L}(z)=r(z)(1+z)
\end{equation}
where $r(z)$ is the comoving angular diameter distance given by Eq.~\ref{eq:rchi}.
We use the latest Union2.1 complation of SNIa dataset reported by Ref.~\citep{2012ApJ...746...85S} 
in the redshift range $0.015\leq z\leq1.414$. This SNIa catalog consists of 580 individual supernova events and is cautiously 
calibrated against numerous sources of systematic uncertainties. 
Compared with the former Union2 compilation \citep{2010ApJ...716..712A}, 
this updated dataset includes twenty-three new events 
at high redshifts ($0.6<z<1.4$) and thus are helpful in 
tightening the constraints on the early behavior of dark energy.

Besides the Union2.1 compilation, there are a number of popular SNIa samples, e.g. ESSENCE 
\cite{2007ApJ...666..694W}, Constitution \cite{2009ApJ...700.1097H}, SDSS-II 
\cite{2009ApJS..185...32K}, SNLS3 \cite{2011ApJS..192....1C}. In deriving those datasets, 
different light curve fitters are adopted. Generally speaking, three independent fitters are 
proposed and maintained amongst the cosmology community, i.e. SALT2 \cite{2007A&A...466...11G}, 
MLCS2k2 \cite{2007ApJ...659..122J}, and SiFTO \cite{2008ApJ...681..482C}. 
When trained on the same SNIa data, SALT2 and SiFTO lead to similar cosmological results. 
However, using the same dataset including SDSS-II, Ref.~\cite{2009ApJS..185...32K} discovered 
a difference of $0.2$ in the best-fits of $w$ by SALT2 and MLCS2k2, which already
exceeds the total error budget (statistical and systematic) contributed from other sources
(also see e.g. Ref.~\cite{2011PhLB..696....5B}, where the discrepancy reaches 3$\sigma$). 
Furthermore, it is found that irrespective of which fitter is selected excluding MLCS2k2, 
different SNIa samples give largely consistent results, however as long as MLCS2k2 
is considered, different datasets lead to notably different constraints on $\Om$ 
\cite{2012ApJ...744..176L}. The major difference between the MLCS2k2 fitter and the other two 
originates from the difference in the rest-frame U band region, where the
training of MLCS2k2 relys exclusively on nearby SNIa observations while the other two also consider
high redshift data \cite{2010ApJ...716..712A}. In order to focus on the constraints on neutrino 
properties, we decide to use the largest sample insofar (i.e. Union2.1) which is obtained from the currently favored and more widely used method (i.e. SALT2). This also allows comparisons with most of other published results.

\subsection{Baryon Acoustic Oscillations}

As the CMB radiation decouples from the primordial photon-baryon plasma, 
the features of acoustic oscillations are imprinted onto matter clustering as well and 
appear as peaks in the galaxy correlation function with a characteristic comoving separation of 
$100\rm~h/Mpc$ \citep{1970ApJ...162..815P, 1996ApJ...471..542H}.
Baryon acoustic oscillations (BAO) thus supply us with a standard 
ruler to measure distances out to the redshift where the bulk of galaxies are observed.

The first detection was achieved by Ref.~\citep{2005ApJ...633..560E}, using SDSS DR3 LRG sample with effective redshift $z=0.35$.
They defined the acoustic parameter which is independent of dark energy models as,
\begin{equation} \label{eq:A}
A\equiv D_{\rm V}(z)\frac{\sqrt{\Om H^2_0}}{zc},
\end{equation}
where the distance combination is depicted by
\begin{equation}
D_{\rm V}(z)=\[r^2(z)\frac{zc}{H(z)}\]^{1/3},
\end{equation}
with $r(z)$ given by Eq.~\ref{eq:rchi}.
We use the BAO measurements from the surveys of SDSS DR7 \citep{2010MNRAS.401.2148P} and WiggleZ \citep{2011MNRAS.418.1707B}.
Hence in total BAO results at five redshifts, i.e. $z=0.2,~0.35,~0.44,~0.6,~0.73$, are 
involved in the constraints. Using the earlier SDSS result, 
Ref.~\citep{2006JCAP...06..019G} suggested that the inclusion of BAO can help break the 
degeneracy between dark energy and neutrino masses.

\subsection{Observational Hubble Parameter Data}

The differential-age technique directly measures the Hubble parameter as 
\begin{equation} \label{hz}
H(z)=-\frac{1}{1+z}\frac{{\rm d}z}{{\rm d}t},
\end{equation}
and therefore endowed the OHD with special functionality as standard clocks.
The quantity 
${{\rm d}z}/{{\rm d}t}$ is usually determined by measuring the age 
differences between passively evolving galaxies with nearly the same spectroscopic redshift \citep{2010JCAP...02..008S}.
We use the most up-to-date datasets summarized by \citep{2012JCAP...07..053M},
which comprises nineteen OHD measurements over the redshift interval from 
$0.09\leq z\leq1.75$.

\subsection{Further Data and Priors}

Complementary to late-time large scale probes (including WL) and other standard geometrical indicators (e.g. SNIa, BAO, OHD), the observation of CMB anisotropies is vital in this analysis. 
We use four independent power spectra, i.e. the temperature auto-power $C^{\rm TT}_l$,
the curl-free component polarization auto-power $C^{\rm EE}_l$, the curl component polarization auto-power $C^{\rm BB}_l$,
and the T-E cross-correlation $C^{\rm TE}_l$, at the last scattering surface 
during the decoupling epoch ($z\sim1090$).
We use the WMAP seven-year data via including the standard pipeline for computing the likelihood,
supplied and maintained by the WMAP collaboration 
\footnote{available at the LAMBDA website, \url{http://lambda.gsfc.nasa.gov/}}. 
Apart from that, a top-hat prior of $10~\textrm{Gyr}<t_0<20~\textrm{Gyr}$ on the cosmic age is employed.
Last but not least, in all data combinations, we always impose the HST prior 
on the Hubble constant as $H_0=74.2\pm3.6\rm~km/s/Mpc$ \citep{2009ApJ...699..539R}.

\section{Constraints on Cosmological Parameters}\label{sec:constrain}

The cosmological parameters we use are presented in Table~\ref{table:priors}. 
Our most generalized parameter space is composed of the ``Vanilla+Extended'' sets of parameters,
represented by the following vector,
\begin{equation}
\textbf{P}\equiv\(\Obh,\Och,\thA,\tau,\ns,\ln{(10^{10} \As)},\{\Sigma m_{\nu},\neff,w\}\).
\end{equation}
In addition, the pivot value of the primordial power spectrum is taken to be $k_0=0.05\rm~Mpc^{-1}$. 
For simplicity, we assume a flat geometry and purely adiabatic initial conditions. 
We note that a non-zero curvature might loosen the constraint on $\mnu$ \citep{2012PhRvD..85l3521S},
but the WMAP seven-year result strongly confines $-0.00133<\Ok<0.0084$ (95\% CL),
which justifies our assumption. Moreover, we do not consider a running spectral index or any tensor contribution,
for the constraint on neutrino masses is not relaxed even when they are allowed \citep{2010JCAP...01..003R}.
We also do not consider an evolving equation of state.

Our code is a modified version of the publicly available package CosmoMC \citep{2002PhRvD..66j3511L}
\footnote{\url{http://cosmologist.info/cosmomc}}. 
The global fitting to parameter vector $\textbf{P}$ is achieved through the exploration of 
the corresponding multidimensional parameter space with the Markov Chain Monte Carlo (MCMC) technique.
In realizing the MCMC approach, the Metropolis-Hastings algorithm is implemented to generate sets 
of chains containing sample points distributed according to the overall likelihood in the parameter space
with top-hat prior probability distribution on each input parameter (see Table~\ref{table:priors}).

Aiming at joint analyses of cosmological probes, we select five combinations from the aforementioned datasets 
(see Sect.~\ref{sec:data}), i.e. CMB, CMB+WL, CMB+BAO+OHD, CMB+BAO+SNIa, and CMB+WL+BAO+OHD+SNIa, 
the last of which is often quoted as the full combination hereafter.
For all data combinations, the HST prior is always used.
For a progressive and intensive exploration on the extended parameter set, 
there exist six scenarios for parameter combinations, i.e.
Vanilla+$\mnu$, Vanilla+$\neff$, Vanilla+$\mnu$+$\neff$, Vanilla+$w$+$\mnu$, Vanilla+$w$+$\neff$, Vanilla+$w$+$\mnu$+$\neff$.
As a consequence, we have attempted more than thirty runs in total, so as to sample the likelihood distribution 
in multivariate space for different combinations of observational data.
For each run, eight chains are simultaneously generated using parallel computation, and they 
stop as soon as the criterion of Gelman and Rubin test ($R$ statistics) is satisfied. 
Generally speaking, after 
convergence, our chains contain $10^5$ points each and guarantee $R-1<0.01$. 
Then these chains are thinned and joined, left with more than 20000 points for the final 
constraints on each scenario for each data combination. The details of our results are reported as follows.

\subsection{\boldmath Vanilla+$\mnu$}

\begin{figure}
    \centering
    \includegraphics[width=.52\textwidth]{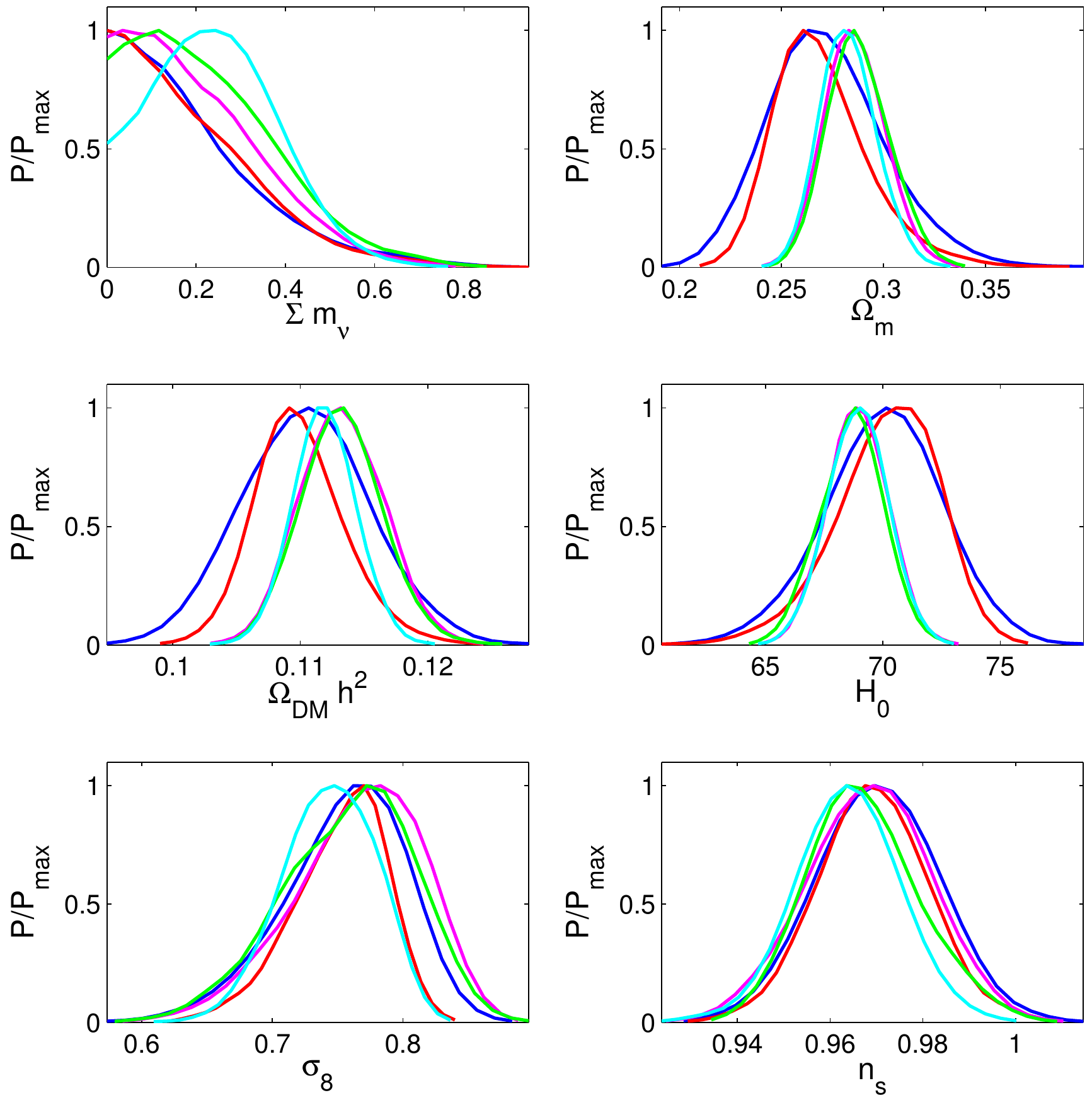}
    \caption{One-dimensional marginalized constraints on individual cosmological parameters
from five combinations of different observations, i.e. CMB (blue), CMB+WL (red),
CMB+BAO+OHD (magenta), CMB+BAO+SNIa (green), and CMB+WL+BAO+OHD+SNIa (cyan). Here
$\mnu$ is free whereas $\neff$ and $w$ are fixed to their standard values 
($\neff=3.046,~w=-1$).}
    \label{fig:lam_mnu}
\end{figure}

\begin{figure}
    \centering
    \includegraphics[width=.6\textwidth]{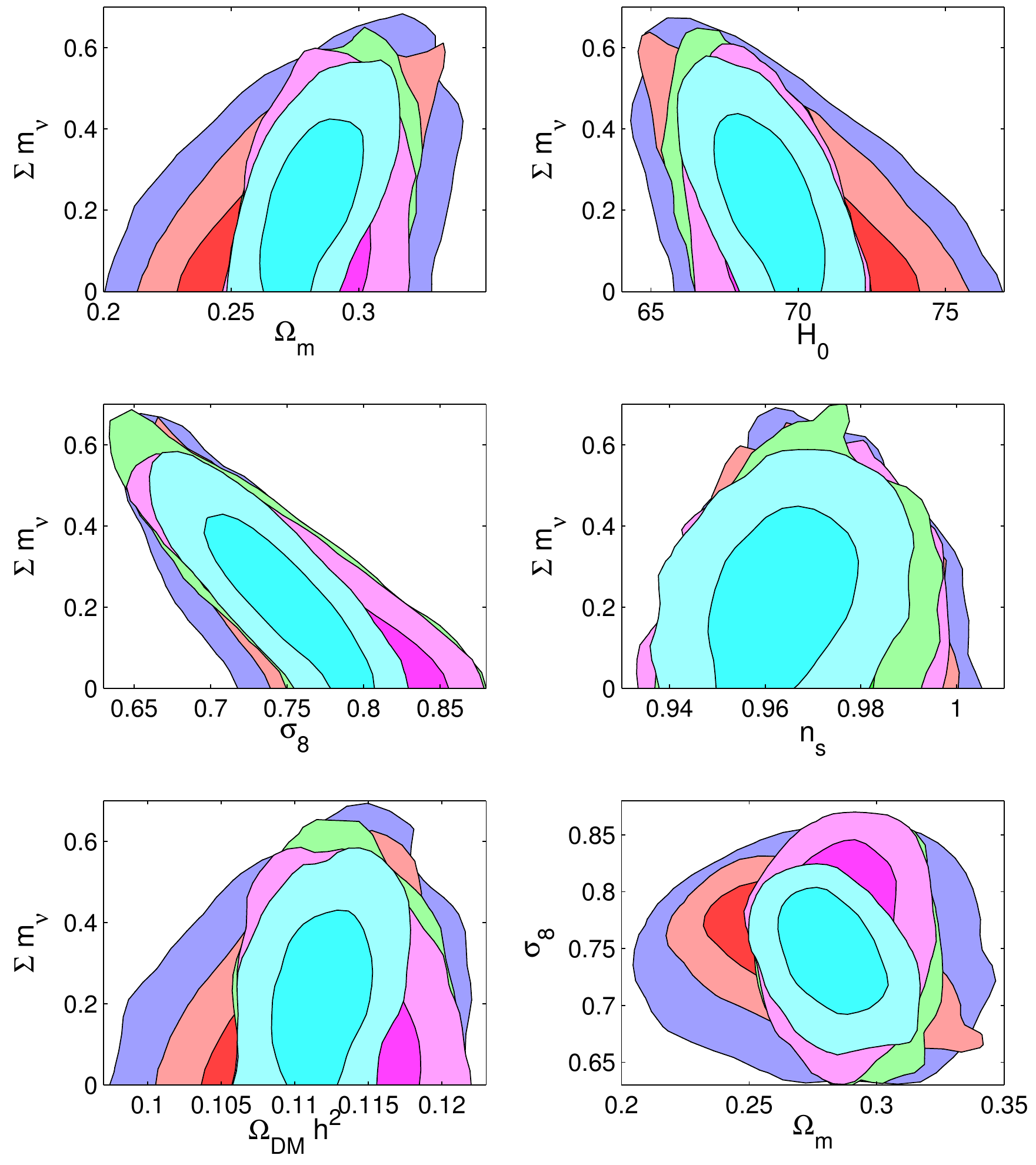}
    \caption{Two-dimensional confidence contours of several parameter pairs  
from five combinations of different observations, i.e. CMB (blue), CMB+WL (red),
CMB+BAO+OHD (magenta), CMB+BAO+SNIa (green), and CMB+WL+BAO+OHD+SNIa (cyan). For each
data combination, the inner and outer contours refer to the borders of 68\% and 95\% confidence 
regions, respectively. As in Fig.~\ref{fig:lam_mnu}, $\mnu$ is free here, but 
$\neff$ and $w$ are fixed.}
    \label{fig:lam_mnu_2D}
\end{figure}

We first explore the parameter space of Vanilla+$\mnu$. 
Our main constraint results are quantitively summarized in Table~\ref{table:lam_mnu}.
Fig.~\ref{fig:lam_mnu} displays the one-dimensional marginalized posterior distribution on some important parameters, while
Fig.~\ref{fig:lam_mnu_2D} shows the two-dimensional confidence contours revealing some crucial parameter degeneracies.
Here as usual, we assume total degeneracy for the three basic neutrino eigenstates.
The effective number of neutrinos is fixed as its standard value, i.e. $\neff=3.046$.
The constraints are drawn under the context of the popular cosmological constant model ($w=-1$).
Like many other papers on constraining $\mnu$ while fixing $\neff$ and $w$,
we found a 95\% CL upper bound on the sum of neutrino masses, $\mnu<0.476\rm~eV$,
using the full combination. (Hereafter we neglect the unit of ``eV'' for brevity.) 
This is a 10\% improvement over the formerly reported 
upper limit given by Ref.~\citep{2009A&A...500..657T} and Ref.~\citep{2009PhRvD..79b3520I} using similar data.
The WL evidently contributes to the overall constraints, pushing forward from $\mnu<0.524$ (CMB) 
to $\mnu<0.496$ (CMB+WL), even better than the result given by CMB+BAO+SNIa.
It reveals one of the greatest advantages of the WL technique that it measures the 
matter fluctuation amplitude very accurately, as seen from Table~\ref{table:lam_mnu};
$\sigma_8$ is pinned down to the narrowest 68\% confidence range for CMB+WL.
Moreover, the OHD give similar, even slightly better results than the SNIa do ($\mnu<0.486$ vs. $\mnu<0.518$),
considering the vast difference between the statistical size of the two data sample (19 vs. 580).
This may be persuasive in designing next generation spectroscopic surveys 
in pursuit of large sample OHD measurements \citep{2009MPLA...24.1699L, 2011ApJ...730...74M}.
From Fig.~\ref{fig:lam_mnu_2D}, we see a mild proportionality between $\mnu$ and $\Om$ 
while a slight inverse proportionality between $\mnu$ and $H_0$ is also seen. 
The inverse proportionality between $\mnu$ and $\sigma_8$ 
should be counted as the strongest correlation. Those degeneracies naturally 
explain why our best-fit values for $\Om$, $H_0$ and $\sigma_8$ are mildly larger,
a little smaller, and unambiguously smaller than the WMAP seven-year recommended values.
Except for the three derived parameters, constraints on the Vanilla parameter set retrieve nearly the same results
as reported by the WMAP seven-year analysis.
In particular, including additional probes does not result in much better constraints on $\ns$ 
and the degeneracy between $\mnu$ and $\ns$ is marginal, since the variation of 
$\ns$ changes the global shape of matter power spectrum whereas $\mnu$ only damps $P(k)$ at small scales.

\subsection{\boldmath Vanilla+$\neff$}

\begin{figure}
    \centering
    \includegraphics[width=.52\textwidth]{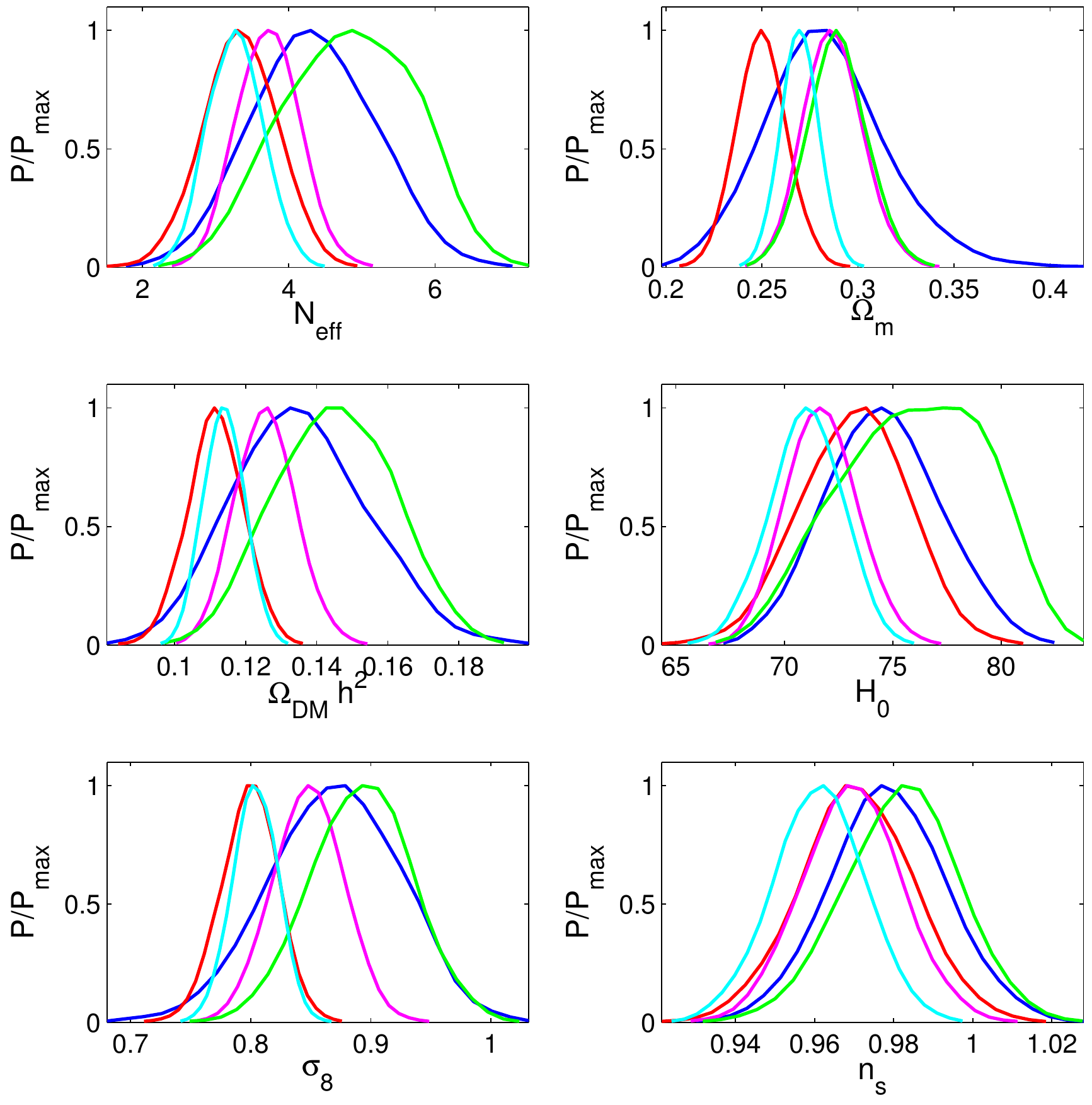}
    \caption{Same as Fig.~\ref{fig:lam_mnu}, except that $\neff$ is a free parameter
here, with massless neutrinos and $w=-1$.}
    \label{fig:lam_Neff}
\end{figure}

\begin{figure}
    \centering
    \includegraphics[width=.6\textwidth]{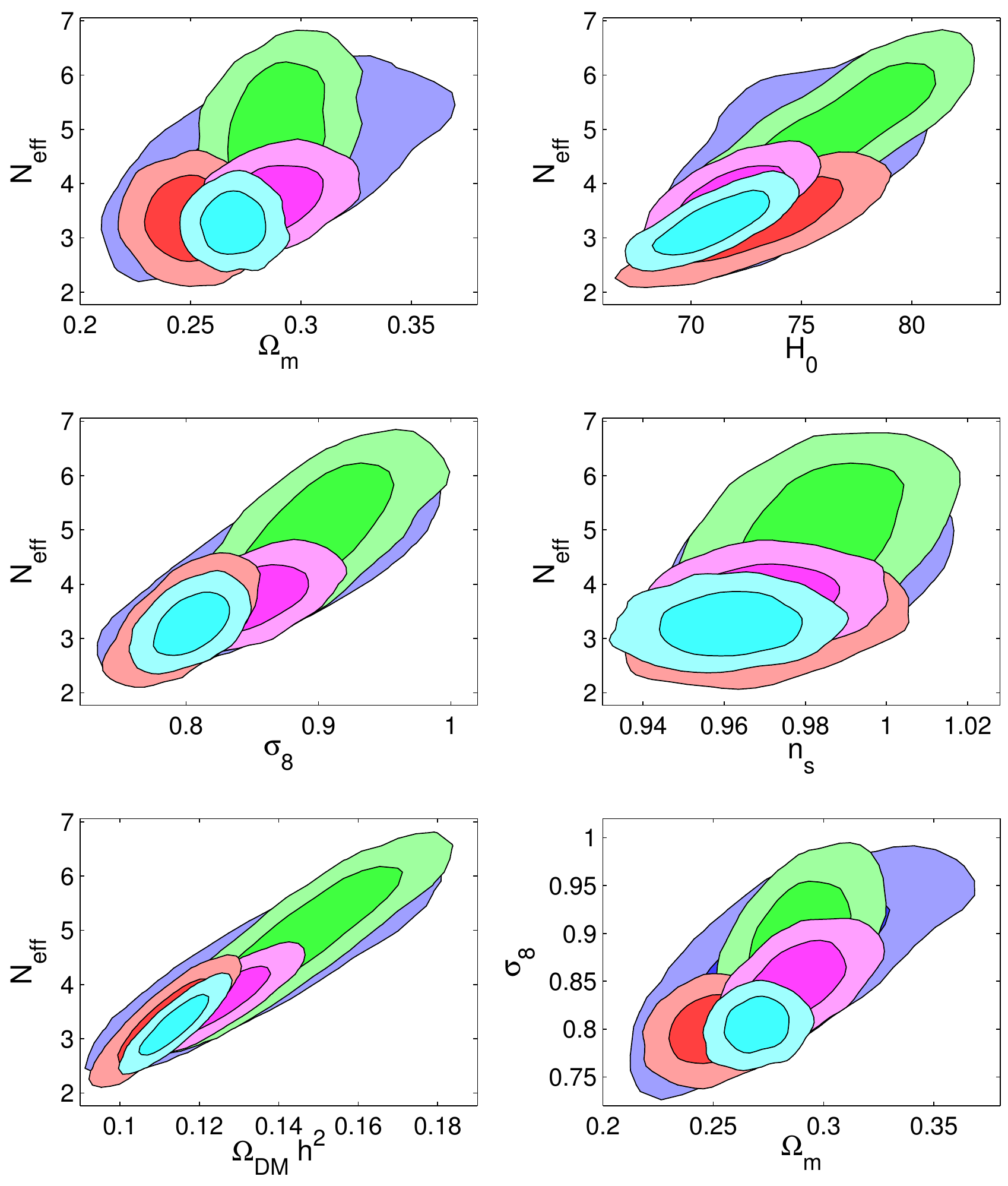}
    \caption{Same as Fig.~\ref{fig:lam_mnu_2D}, except that $\neff$ is a free parameter
here, with massless neutrinos and $w=-1$.}
    \label{fig:lam_Neff_2D}
\end{figure}

Then we test the scenario of Vanilla+$\neff$ against each cosmological data combination. 
Likewise, the main results are presented in Table~\ref{table:lam_Neff}, 
Fig.~\ref{fig:lam_Neff} (one-dimensional) and Fig.~\ref{fig:lam_Neff_2D} (two-dimensional).  
Here the global fitting is completed under the assumption of massless neutrino with 
$\Lambda$CDM cosmology. Using the full combination of cosmological probes, we obtain $\neff=3.271\pm0.367$,
which rules out additional light degrees of freedom at nearly 95\% CL. 
This result is  in support of the standard prediction of big bang nucleosynthesis, 
and is primarily due to the contribution of the WL data 
($\neff=3.334\pm0.496$ for CMB+WL, whereas $\neff=4.340\pm0.817$ for CMB alone). 
The OHD also tend to favor the standard value. Including OHD, we obtain $\neff=3.722\pm0.418$, 
which is fully consistent with Ref.~\citep{2012JCAP...07..053M} ($\neff=3.7\pm0.4$). 
Once again OHD give rather tighter bounds on the effective number of neutrinos than the SNIa do,
and this time the difference between their constraining powers is more clearly revealed.
The SNIa data seem to prefer the largest $\neff$ best-fit value with the largest uncertainty.
In this case, the inclusion of the SNIa data does not improve the constraint on either $H_0$ 
or $\Och$. Fig.~\ref{fig:lam_Neff_2D} displays several parameter correlations.
As reported by Ref.~\citep{2009ApJS..180..306D}, we see the strong anticorrelation between
$\neff$ and $\Och$. The degeneracy between $\neff$ and $H_0$ is also remarkably 
strong. The propotional trend between $\neff$ and $\Om$ is mild and disappears 
when the full combination is employed. 
These correlations arise from the role $\neff$ plays in the radiation content,
which decides the epoch of matter-radiation equality as
\begin{equation} \label{eq:zeq}
1+z_{\textrm{eq}}=\frac{\Om}{\Or}=\frac{\Om h^2}{1+0.2271\neff} (\Omega_{\gamma}h^2)^{-1}.
\end{equation}
CMB measures precisely the present-day photon energy density to be 
$\Omega_{\gamma}h^2=2.469\times10^{-5}$ for $T_{\gamma0}=2.725~\rm K$ \citep{2011ApJS..192...18K}.
Given the rather narrow range of $z_{\textrm{eq}}=3141^{+154}_{-157}$ \citep{2009ApJS..180..330K},
Eq.~\ref{eq:zeq} thereby explains the correlations observed between
$\neff$, $\Om$, and $H_0$. The degeneracy between $\neff$ and $\Och$ is also extracted 
from Eq.~\ref{eq:zeq}, since $\Om h^2=\Obh+\Och$ and meanwhile
the physical baryon density $\Obh$ is tightly constrained by the CMB observations.
Unlike its inverse proportionality with $\mnu$, $\sigma_8$ shows clear proportionality with 
$\neff$. Because the observation of WL provides an accurate and independent measurement on $\sigma_8$,
we thus expect that WL plays the most influential role in current constraints.
The inclusion of SNIa is inadequate to deal with parameter degeneracies associated with $\neff$,
and that is the reason for some of the poor constraints given by CMB+BAO+SNIa. 
Like the scenario of Vanilla+$\mnu$, the constraints on the Vanilla parameter set 
are similar to the WMAP seven-year results.
and the degeneracy between $\neff$ and $\ns$ is broken when all data are included.

\subsection{\boldmath Vanilla+$\mnu$+$\neff$}\label{subsec:l_mN}

\begin{figure}
    \centering
    \includegraphics[width=.52\textwidth]{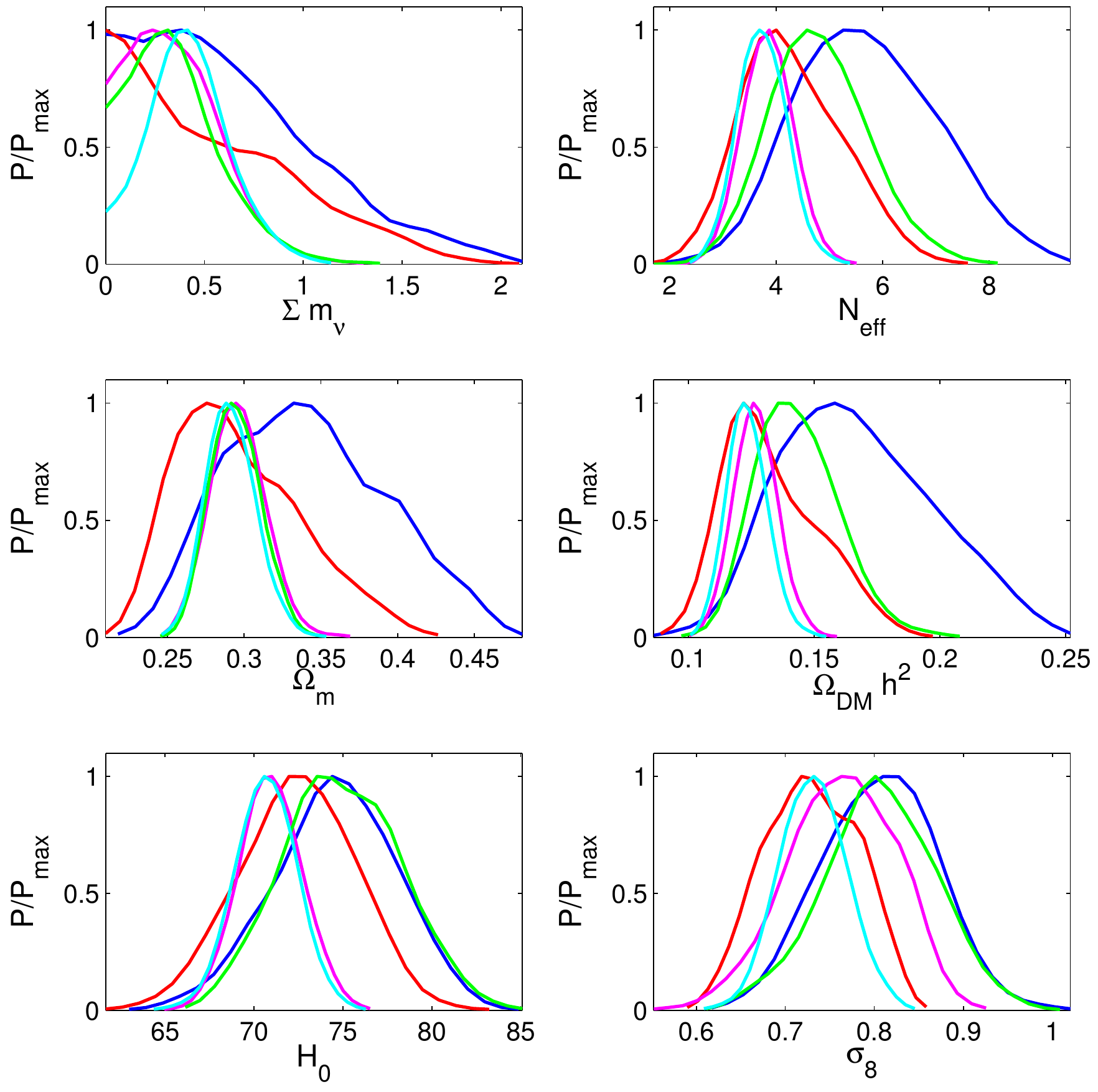}
    \caption{Same as Fig.~\ref{fig:lam_mnu}, but for both $\mnu$ and $\neff$ kept free, with $w=-1$.}
    \label{fig:lam_mnuNeff}
\end{figure}

\begin{figure}
    \centering
    \includegraphics[width=.75\textwidth]{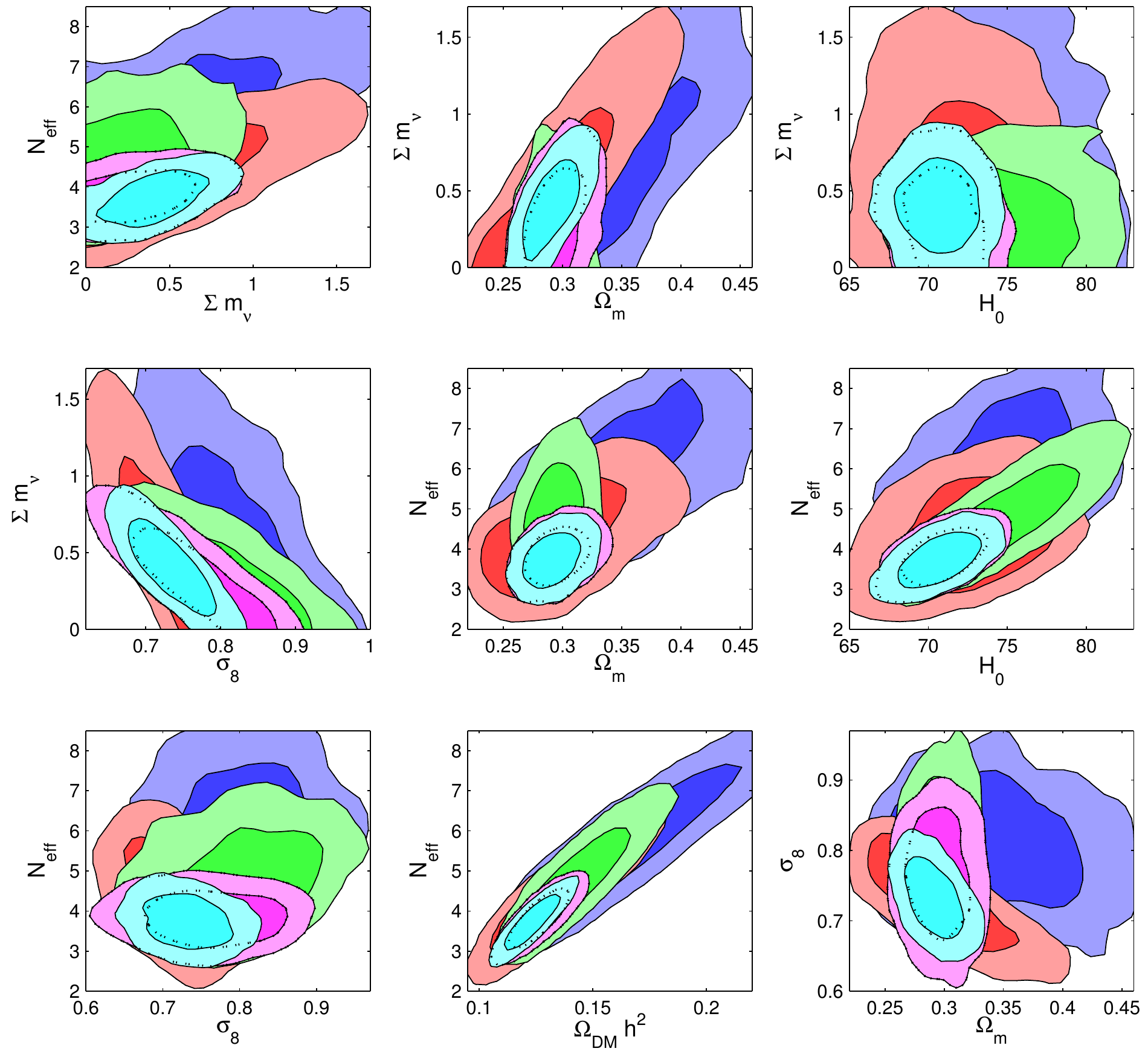}
    \caption{Same as Fig.~\ref{fig:lam_mnu_2D}, but for both $\mnu$ and $\neff$ kept free, 
with $w=-1$. For the sake of clarity, the dotted lines in each panel 
denote the same confidence regions for CMB+BAO+OHD, as shown by the magenta 
contours.}
    \label{fig:lam_mnuNeff_2D}
\end{figure}

We next investigate the case when both $\mnu$ and $\neff$ are set free, 
under the assumption of cosmological constant. Our main constraints are presented 
in Table~\ref{table:lam_mnuNeff}, Figs.~\ref{fig:lam_mnuNeff} and \ref{fig:lam_mnuNeff_2D}. 
Undoubtedly, the inclusion of $\neff$ as a free parameter tremendously weakens the constraints
on $\mnu$. Interestingly, when the full combination is considered, a robust 68\% lower limit
is revealed and our result is $\mnu=0.421^{+0.186}_{-0.219}$ at 68\% CL.
Nonetheless, considering that the best-fit of our current result is almost excluded at 95\% CL by
our former constraint when $\neff$ is fixed ($\mnu<0.476$), this may illustrate our worry 
on the robustness of all stringent upper bounds on $\mnu$ ever reported in previous literature.
The constraints on $\neff$ are also relaxed due to the freedom on $\mnu$.
Unlike former results, the extra neutrino species is slightly favored, as our best constraint reads $\neff=3.740\pm0.446$.
The WL data still shows great promise as it refines the constraints from ($\mnu<1.515,~\neff=5.729\pm1.274$, CMB alone)
to ($\mnu<1.393,~\neff=4.308\pm0.924$, CMB+WL), mainly due to its precise constraint on $\sigma_8$.
The combination of CMB+BAO+OHD(+HST) (Fig.~\ref{fig:lam_mnuNeff_2D}) exerts more dominant influence on the parameter constraints, 
as the cyan and magenta contours are more alike. 
Besides the enormous information from the WMAP seven-year data, 
the OHD+HST delivers direct knowledge on the expansion history. 
In combination with the BAO distance indicator, this combination therefore is capable of 
successfully breaking several hard parameter degeneracies, e.g. ($\mnu$ vs. $H_0$), 
($\neff$ vs. $\Om$). However even against the full combination, 
some correlations of parameters still remain. For instance, 
a larger $\mnu$ value shows a preference for a smaller $\sigma_8$ yet a slightly larger $\Om$,
whereas a larger $\neff$ favors a larger $\Och$ and/or $H_0$. 
Most importantly, the mild proportionality between ($\mnu$, $\neff$) reasonably points 
to the fact that a larger total neutrino mass is preferred by a larger effective number. 
As an increasing number of observations infer extra light degrees of freedom, 
varying $\mnu$ and $\neff$ simultaneously will be a non-trivial 
consideration to derive constraints on both $\mnu$ and $\neff$.
As in previous scenarios, the constraints on the Vanilla parameter set are mostly unaffected. 

\subsection{\boldmath Vanilla+$w$+$\mnu$}

\begin{figure}
    \centering
    \includegraphics[width=.52\textwidth]{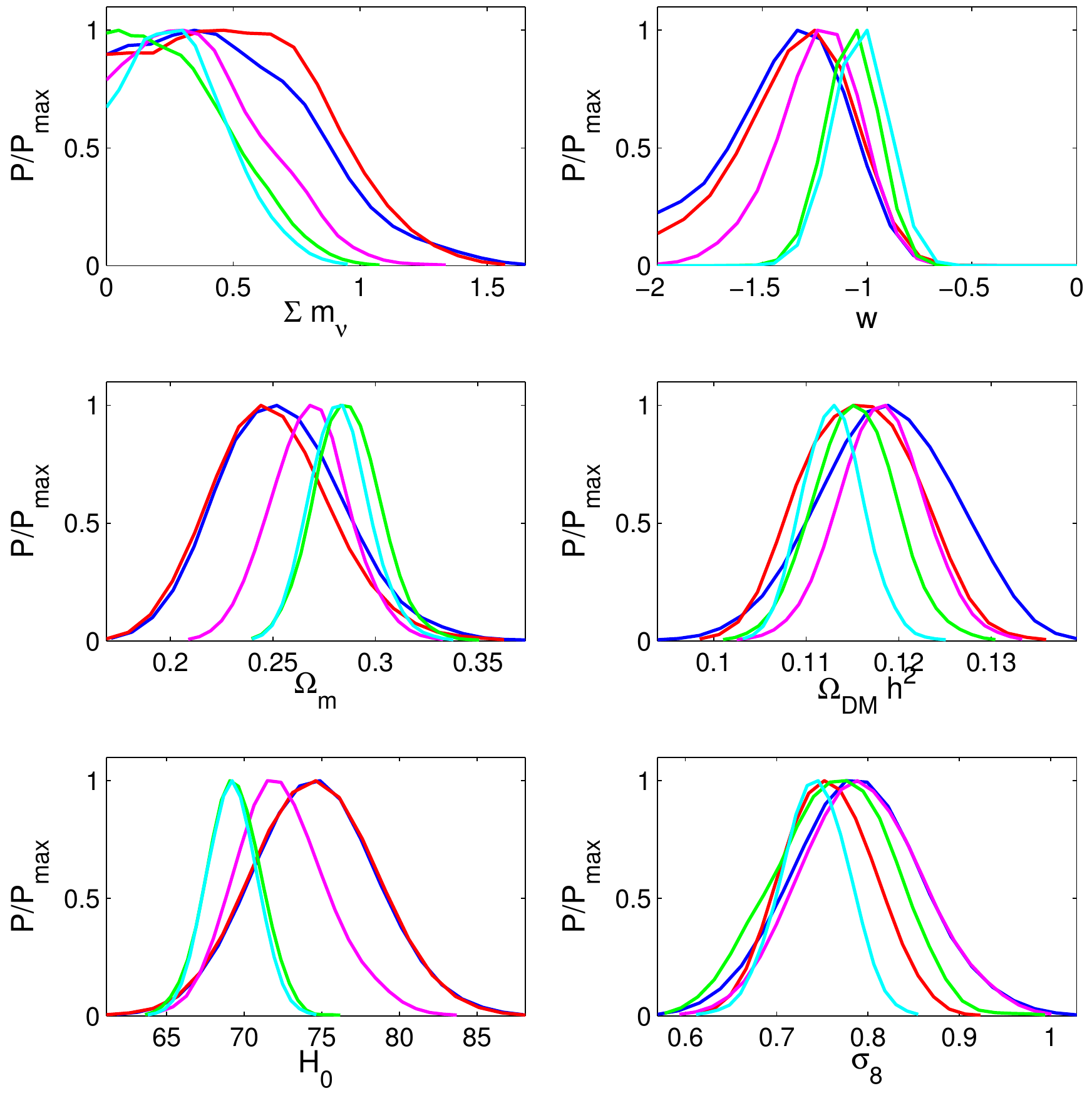}
    \caption{Same as Fig.~\ref{fig:lam_mnu}, but also with $w$ freed, and $\neff$ is still fixed.}
    \label{fig:w_mnu}
\end{figure}

\begin{figure}
    \centering
    \includegraphics[width=.6\textwidth]{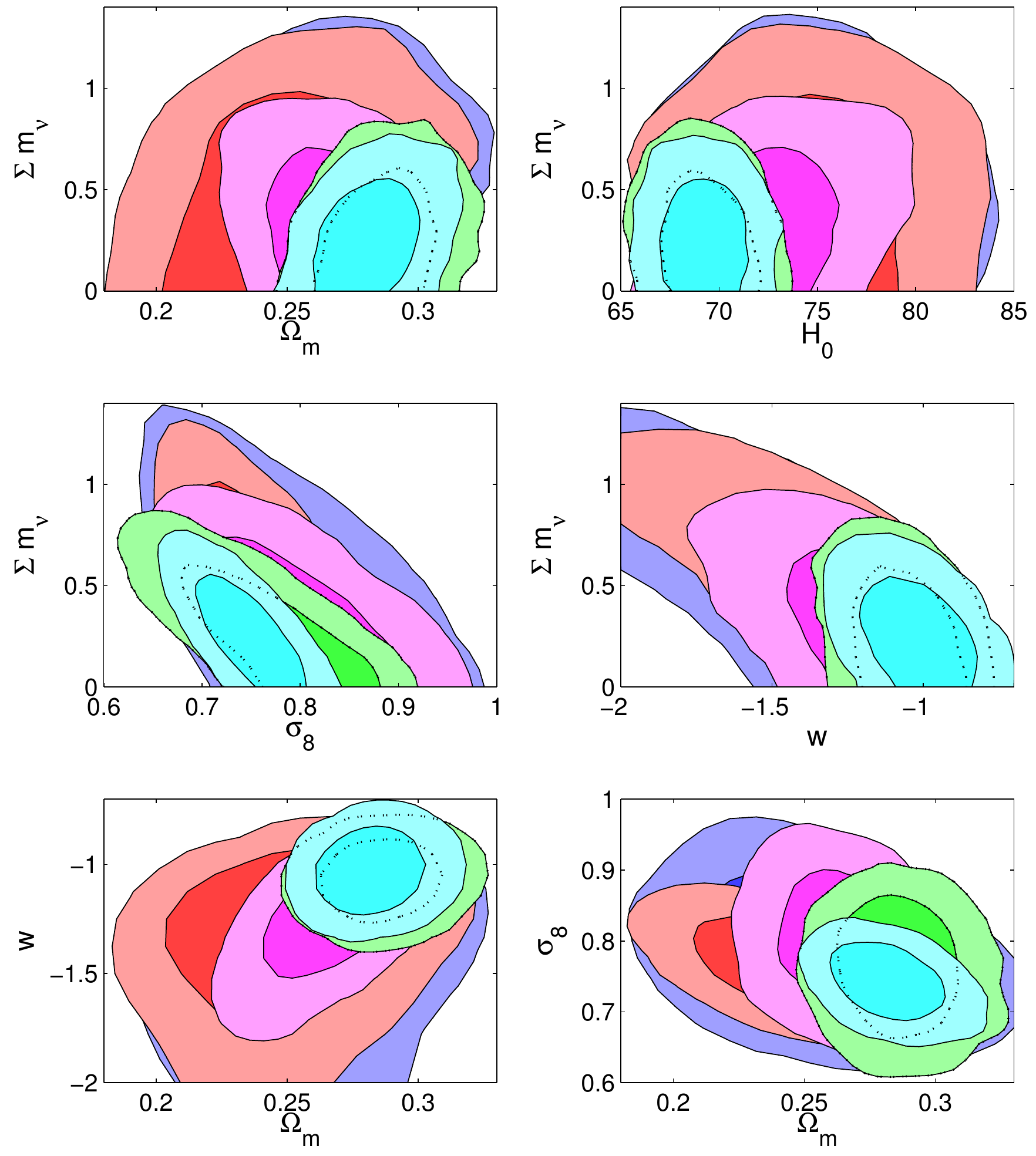}
    \caption{Same as Fig.~\ref{fig:lam_mnu_2D}, but also with $w$ freed, 
and $\neff$ is still fixed. For the sake of clarity, the dotted lines in each panel
denote the same confidence regions for CMB+BAO+SNIa, as shown by the green contours.}
    \label{fig:w_mnu_2D}
\end{figure}

The previous stage of our analysis is constructed under the hypothesis of the cosmological constant
being the underlying dark energy model. In what follows, we shall assume the dark energy component has a
variable EOS, $w$. 
The other free extended parameter is $\mnu$. The main results for this scenario are  
shown in Table~\ref{table:w_mnu}, Figs.~\ref{fig:w_mnu} and \ref{fig:w_mnu_2D}. 
Owing to the inclusion of $w$ as a free parameter, the best constraint on the sum of neutrino masses
relaxes to $\mnu<0.627$ (with no lower limit observed). 
In this scenario, the primary predominance the SNIa data holds over the OHD is demonstrated,
i.e. the special importance of the SNIa data in constraining $w$. 
In practice, the SNIa impose much tighter constraint ($w=-1.074\pm0.088$) than the OHD
($w=-1.240\pm0.182$). Due to the degeneracy between $w$ and $\mnu$,
CMB+BAO+SNIa thus presents better result ($\mnu<0.688$) than CMB+BAO+OHD gives ($\mnu<0.819$).
Moreover, the combination of CMB+BAO+SNIa furnish an overally strigent constraint, 
as the green contours almost overlap with the cyan ones shown in Fig.~\ref{fig:w_mnu_2D}.
Besides, adding the present WL data does not help much at improving the constraints when $w$ is allowed to vary freely.
Although the technique of WL has been forecasted to be a promising tool to discriminate between dark energy and neutrino masses 
\citep{2003PhRvL..91d1301A}, currently speaking the WL data do not provide this discerning ability, 
albeit its constraint on $\sigma_8$ is still incomparable.
The anticorrelation between $\mnu$ and $\sigma_8$ is strong as usual, 
while $\mnu$ and $w$ are found to be highly
correlated. Nevertheless, the degeneracies between $\mnu$, $\Om$ and $H_0$ are mild.  
Especially, the inclusion of the SNIa data manages to break the degeneracy of ($\Om$ vs. $w$).

\subsection{\boldmath Vanilla+$w$+$\neff$}

\begin{figure}
    \centering
    \includegraphics[width=.52\textwidth]{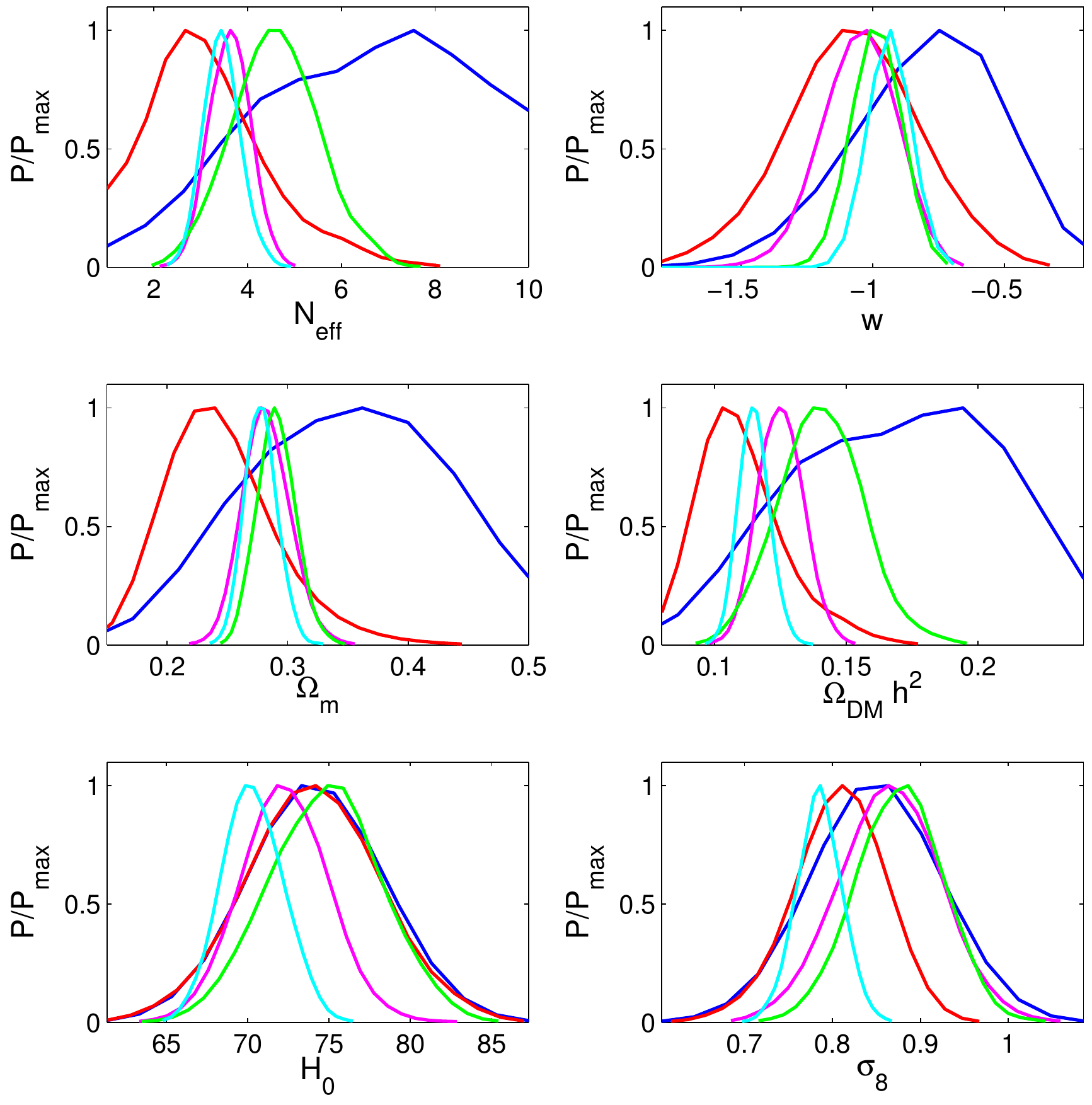}
    \caption{Same as Fig.~\ref{fig:lam_mnu}, but here $\neff$ and $w$ are instead free to vary, 
with massless neutrinos.}
    \label{fig:w_Neff}
\end{figure}

\begin{figure}
    \centering
    \includegraphics[width=.6\textwidth]{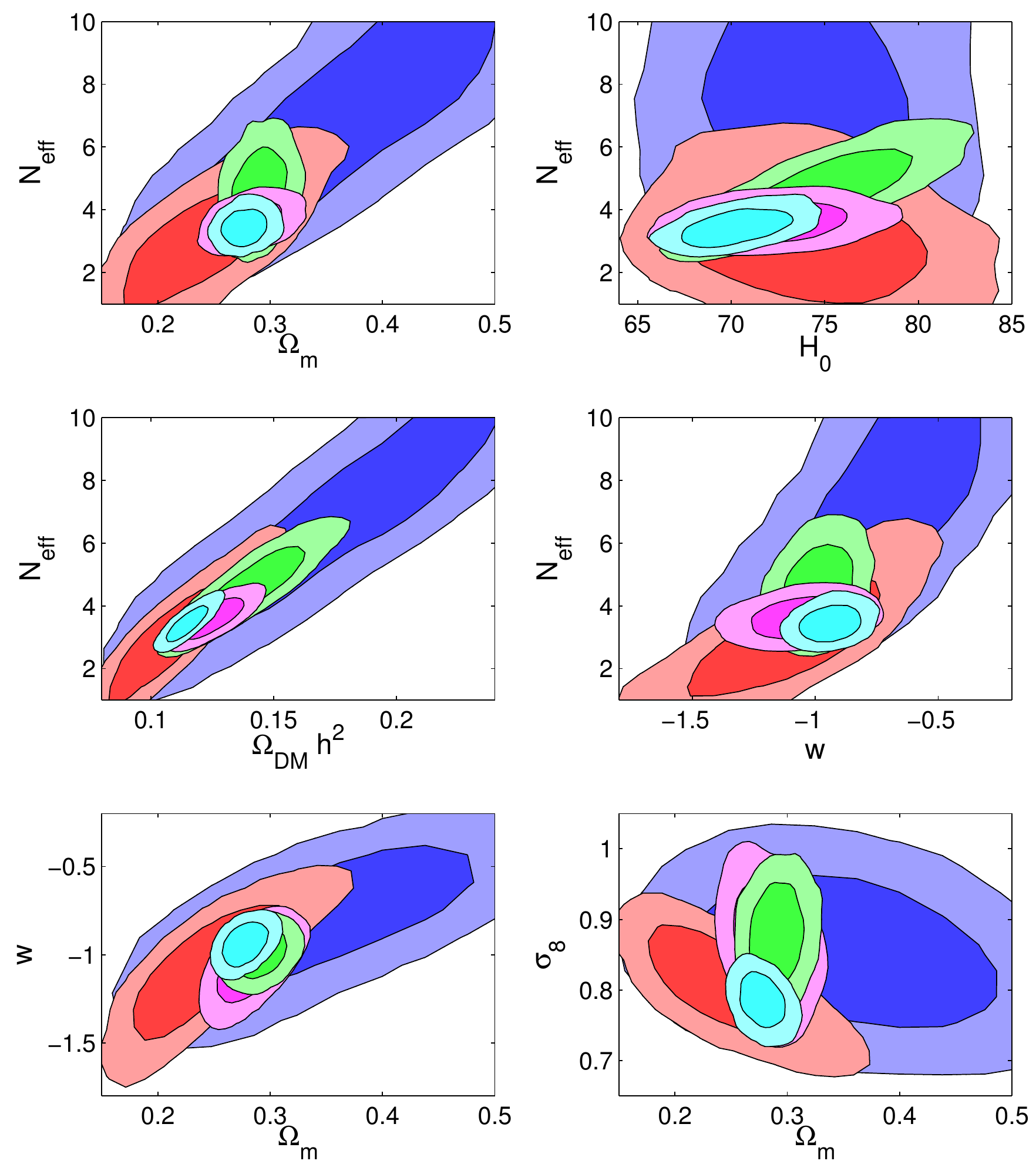}
    \caption{Same as Fig.~\ref{fig:lam_mnu_2D}, but here $\neff$ and $w$ are instead free to vary,
with massless neutrinos.}
    \label{fig:w_Neff_2D}
\end{figure}

The next parameter scenario to explore is Vanilla+$w$+$\neff$.
Table~\ref{table:w_Neff}, Figs.~\ref{fig:w_Neff} and \ref{fig:w_Neff_2D} present the main results.
Unlike the great impact varying $w$ exerts on the $\mnu$ constraints,
the best constraint on $\neff$ is only compromised a little by freeing $w$, which reads $\neff=3.454\pm0.386$.  
But the inclusion of the free EOS parameter does make the combination of CMB(+HST) lose their capability 
to derive a reliable upper limit on $\neff$. 
However the importance of WL reappears unlike the case of Vanilla+$w$+$\mnu$; 
the WL data constrains $\neff=3.192\pm1.214$. Here the OHD wields incomparable 
constraining power on the effective number of neutrino species, i.e. $\neff=3.623\pm0.432$, 
whereas the SNIa luminosity distance measurements still overpredict $\neff$ to an exceedingly large value.
On $w$, the SNIa data without the contributions of WL or OHD strongly constrain $w=-0.986\pm0.077$.
In Fig.~\ref{fig:w_Neff_2D}, interesting features again show up.
The proportionality between $\neff$ and $\Och$ is always strong and impossible to suppress.
The degeneracy between $\neff$ and $\Om$ is broken when the full combination is employed. 
The correlation between $\neff$ and $H_0$ is present, yet weak.
Particularly, the degeneracy directions between $\neff$ and $w$, 
given by CMB+BAO+OHD and CMB+BAO+SNIa, show a nearly orthogonal feature.
It is thereby enlightening since the OHD have the advantage of better constraining $\neff$ while the SNIa are capable 
of pinning down $w$ more precisely. So when the OHD and the SNIa are combined, it is easy to break
the degeneracy between $\neff$ and $w$, as shown in our plots.

\subsection{\boldmath Vanilla+$w$+$\mnu$+$\neff$}\label{subsec:w_mN}

\begin{figure}
    \centering
    \includegraphics[width=.52\textwidth]{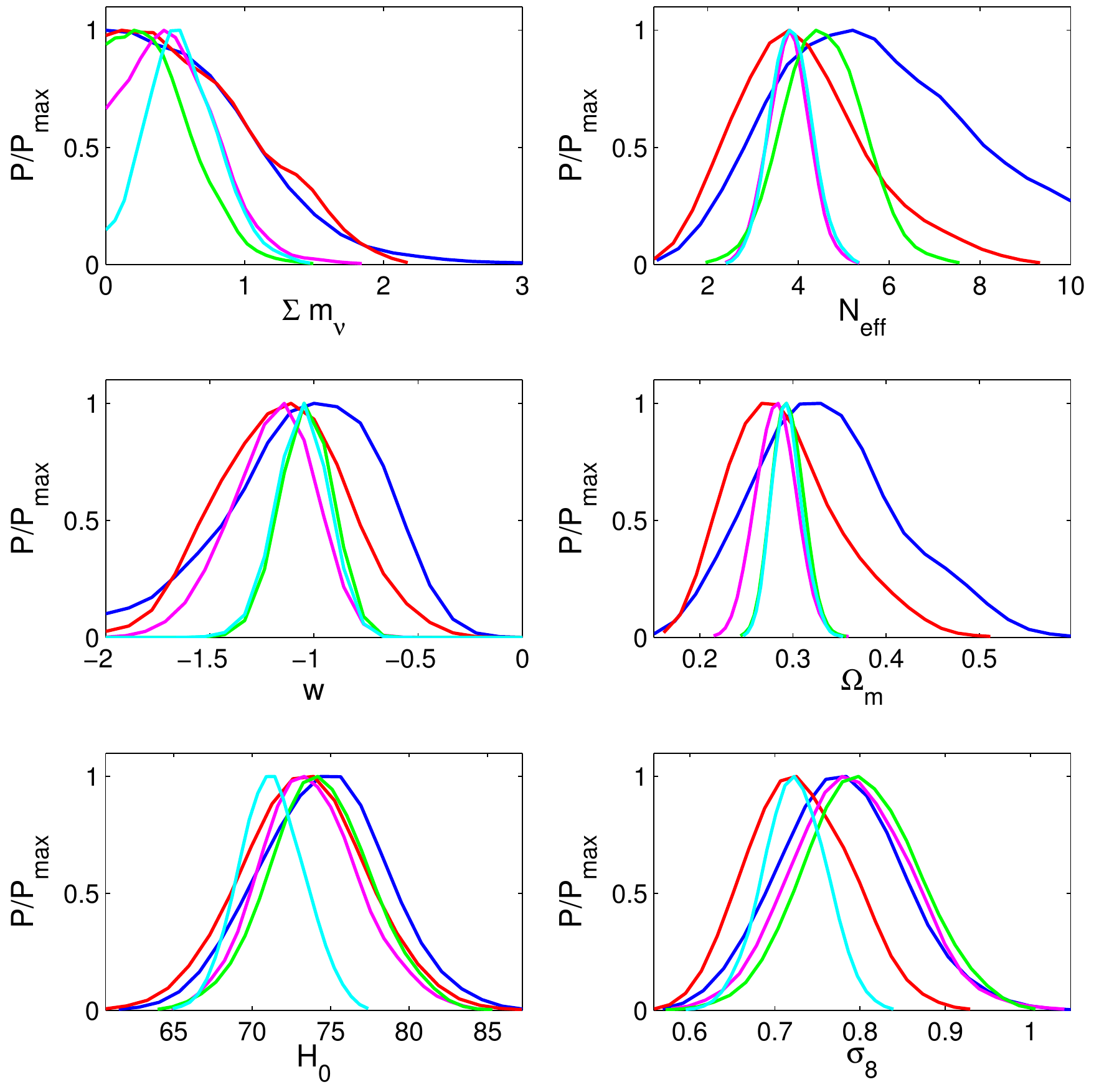}
    \caption{Same as Fig.~\ref{fig:lam_mnu}, yet all members of the extended set 
($\mnu,~\neff,~w$) are treated as free parameters.}
    \label{fig:w_mnuNeff}
\end{figure}

\begin{figure}
    \centering
    \includegraphics[width=.75\textwidth]{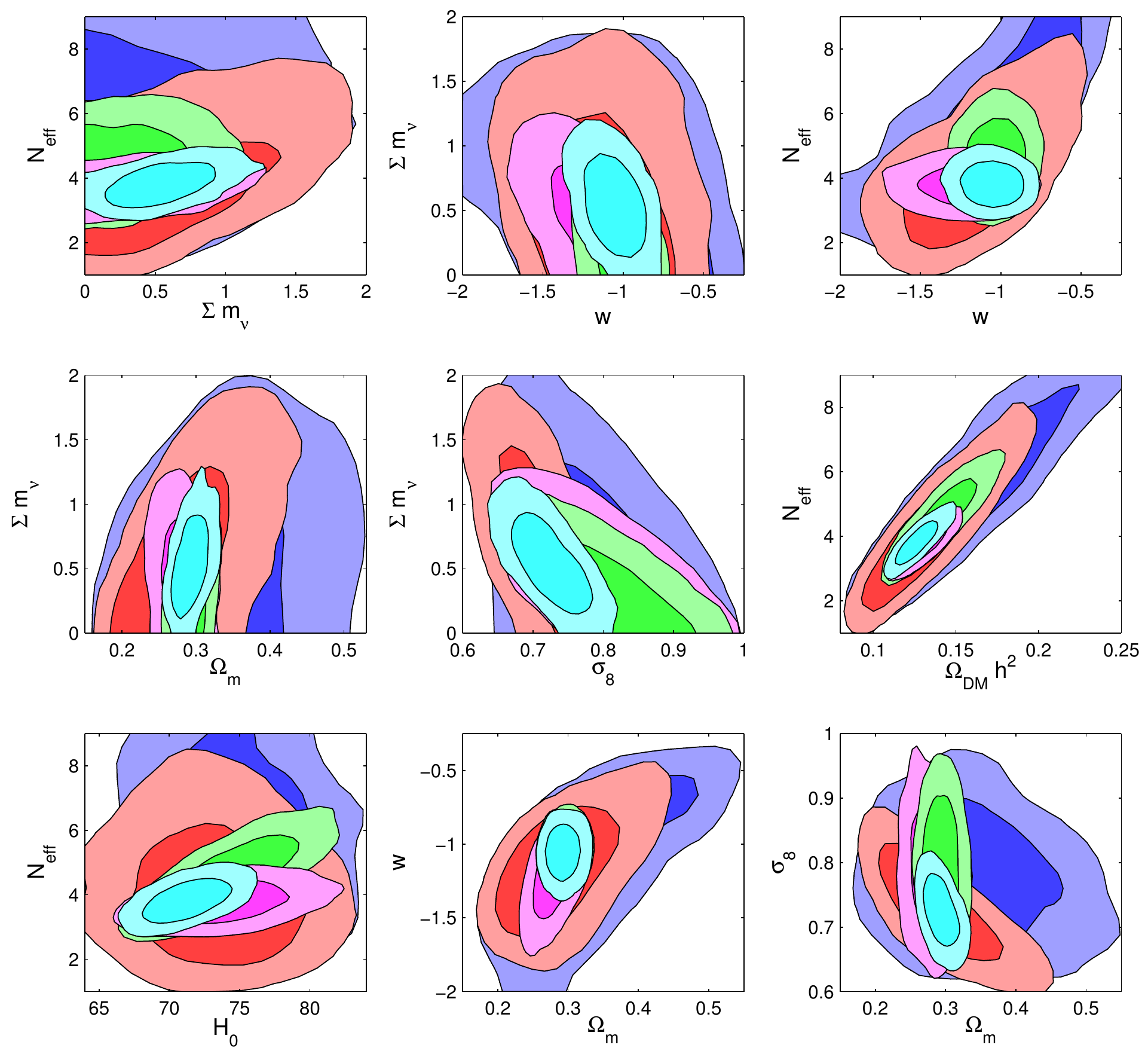}
    \caption{Same as Fig.~\ref{fig:lam_mnu_2D}, yet all members of the extended set
($\mnu,~\neff,~w$) are treated as free parameters.}
    \label{fig:w_mnuNeff_2D}
\end{figure}

Finally, the most generalized scenario, where all of the three extended parameters are free to vary, is analyzed. Our main results are given in Table~\ref{table:w_mnuNeff}, Figs.~\ref{fig:w_mnuNeff} and \ref{fig:w_mnuNeff_2D}.
The best-fits give $\mnu=0.556^{+0.231}_{-0.288}$, $\neff=3.839\pm0.452$, and $w=-1.058\pm0.088$.
Once again a 68\% CL lower limit for $\mnu$ is robustly observed.
As discussed in previous scenarios, the WL data contribute most to the constraint on $\sigma_8$,
and the narrowest 68\% confidence intervals of $w$ and $\Om$ result from the inclusion of the SNIa data.
In addition, due to the degeneracy of ($\mnu$ vs. $w$), the SNIa offer better constraint on $\mnu$ than the OHD. 
However, concerning $\neff$, it is still the OHD that provide the most strict constraint. 
Moreover, it is found that for various scenarios, when $\neff$ is kept free, CMB+BAO+SNIa presents
much weaker constraints on $\Och$ than CMB+BAO+OHD, yet their constraints look similar 
otherwise. This is a straightforward consequence of Eq.~\ref{eq:zeq}. Since CMB+BAO+OHD delivers 
direct information on the expansion history, it provides the most stringent constraint on $\Och$. 
As seen from Fig.~\ref{fig:w_mnuNeff_2D}, when the entire set of extended parameters are freed, 
some of the parameter correlations turn mild, e.g. ($\neff$ vs. $H_0$) and ($\mnu$ vs. $w$). 
Nevertheless, the degeneracies of ($\mnu$ vs. $\sigma_8$) and 
($\neff$ vs. $\Och$) are persistent. 
Once again the confidence contours of ($\neff$ vs. $w$) display remarkable features: CMB+BAO+OHD 
and CMB+BAO+SNIa impose strict constraints with nearly orthogonal degeneracy directions, which 
emphasizes the importance of combining OHD and SNIa in future studies.
Albeit mild, the proportion between $\mnu$ and $\neff$ still exists. It again implies that one
should vary $\mnu$ and $\neff$ simultaneously in order to obtain more reliable constraints on 
cosmic relic neutrino properties.

We note that within most scenarios, the best-fit values of $\sigma_8$ are slightly low
compared with the observations of galaxy clusters or Ly$\alpha$ forest \citep{2005PhRvD..71j3515S, 2012arXiv1203.5775R}.
However we argue that the results are strongly affected by the anticorrelation between $\mnu$ and $\sigma_8$. 
In our trial run with all the extended parameters fixed at their standard values, 
i.e. the usual $\Lambda$CDM scenario, we obtain $\sigma_8=0.791\pm0.018$ and 
$\Om=0.255\pm0.015$, which are in total agreement with Ref.~\citep{2009A&A...497..677K}.

\section{Neutrino Impact on Structure Formation}\label{sec:discus}
Over thirty sets of constraint results have been presented and discussed.
In order to understand the physical origins 
underneath those parameter degeneracies, we investigate 
specifically the influences of the three extended parameters 
on growth of perturbation and formation of structures.
We calculate a variety of matter power spectra via the CAMB code \citep{2000ApJ...538..473L}
\footnote{\url{http://camb.info/}} with the HALOFIT--$\nu$ algorithm \citep{2012MNRAS.420.2551B}.
Here the default parameter set (``ALL'') refers to the best-fit values of the Vanilla+$w$+$\mnu$+$\neff$ scenario
against the full data combination (see Table~\ref{table:w_mnuNeff}). Hereafter, for simplicity, we assume total degeneracy 
for the standard three eigenstates of neutrinos, with mass for each flavor given by $m_{\nu}=\(\mnu\)/3$.

As mentioned in Sect.~\ref{sec:intro}, massive neutrinos influence structure formation 
due to their large thermal velocity dispersion \citep{2006PhRvD..73h3520T}
\begin{equation}
\sigma_{v}(z) = \sqrt{\frac{15\zeta(5)}{\zeta(3)}} \frac{T_{\nu}(z)}{m_{\nu}},
\end{equation}
where $\zeta(s)$ denotes the Riemann zeta function, and $T_{\nu}$ represents the neutrino temperature, which can be 
associated with the present-day temperature for the CMB, i.e. $T_{\nu}(z)=\(4/11\)^{1/3}T_{\gamma0}(1+z)$.
This kinematic activity, i.e. the effect of free-streaming, prevents neutrinos from clustering below the free-streaming scale, 
and thus ensures the density perturbation of neutrinos below this scale negligible.
By analogy with the Jeans length, this characteristic redshift-dependent scale is defined as \citep{1980PhRvL..45.1980B},
\begin{equation}
k_{\rm FS}(z)= \sqrt{\frac{3}{2}}\frac{H(z)}{(1+z)~c_{s,\nu}(z)},
\end{equation}
with $H(z)$ given by Eq.~\ref{eq:expan}, and $c_{s,\nu}$ being the sound speed for thermalized neutrinos,
which can be strictly related to the velocity dispersion through $c_{s,\nu}=\frac{\sqrt{5}}{3}\sigma_{v}$, 
in the non-relativistic limit \citep{2010PhRvD..81l3516S}. 
Neutrinos become non-relativistic when their energy density $\left<E\right>=\frac{7\pi^4}{180\zeta(3)}T_{\nu}=m_{\nu}$,
and this transition occurs at $1+z_{\rm NR}\sim1890(m_{\nu}/1\rm~eV)$. Therefore for masses smaller than $\sim0.6\rm~eV$ 
(the HALOFIT--$\nu$ algorithm is well calibrated within this mass range), neutrinos will not dramaticaly slow down 
until the last scattering of photons ceases. Apart from that, if the neutrinos
are massive enough, i.e. $m_{\nu}>0.005\rm~eV$, to have decelerated before dark energy dominates the expansion history, 
the Hubble horizon at the epoch when they turn into non-relativistic can be described by \citep{2006PhRvD..73h3520T},
\begin{equation}
k_{\rm NR} = \frac{H(z_{\rm NR})}{1+z_{\rm NR}} \sim 0.0145\(\frac{m_{\nu}}{1\rm~eV}\)^{1/2}\Om^{1/2}\rm~h/Mpc.
\end{equation}
This is the largest scale where the presence of finite-mass neutrinos starts to influence low-reshift matter power spectra
by the free-streaming effect; above this scale, massive neutrinos can be treated simply as aggregating CDM.
At scales $k>k_{\rm NR}$, neutrinos not only free-stream to erase their fluctuations, but also damp the amplitude of 
the total matter power spectra by at least a few percent. 

\begin{figure}
    \centering
    \includegraphics[width=.6\textwidth]{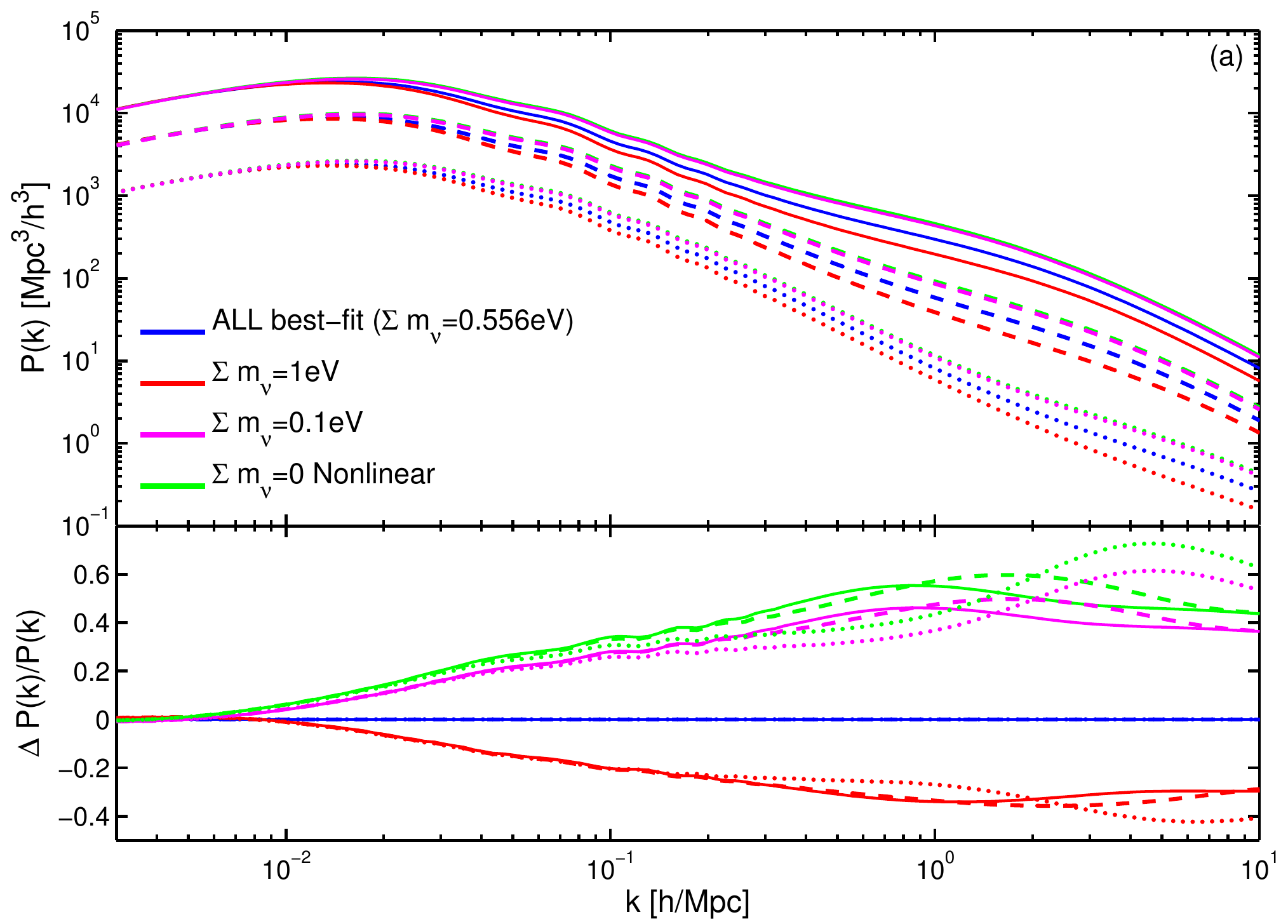}
    \includegraphics[width=.6\textwidth]{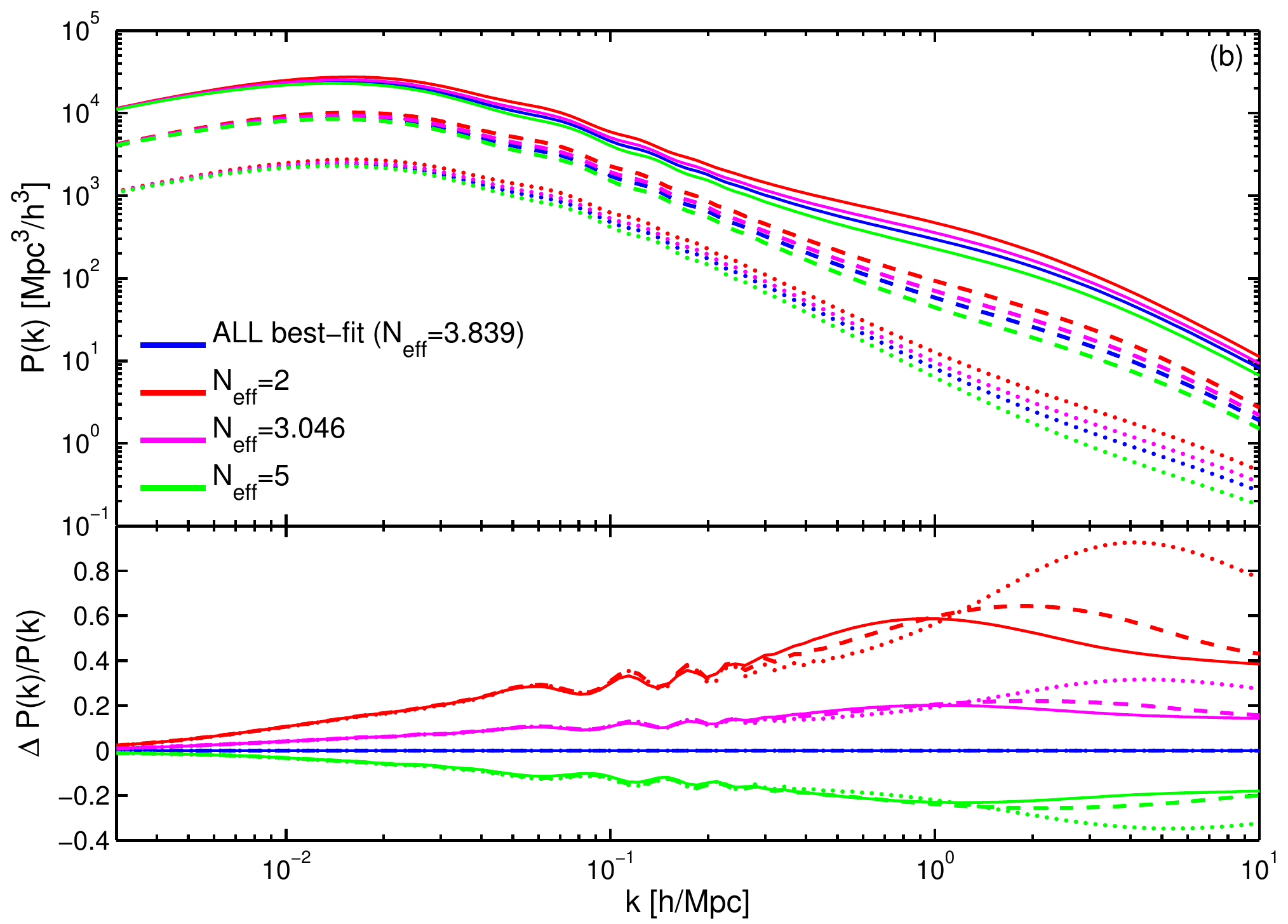}
    \includegraphics[width=.6\textwidth]{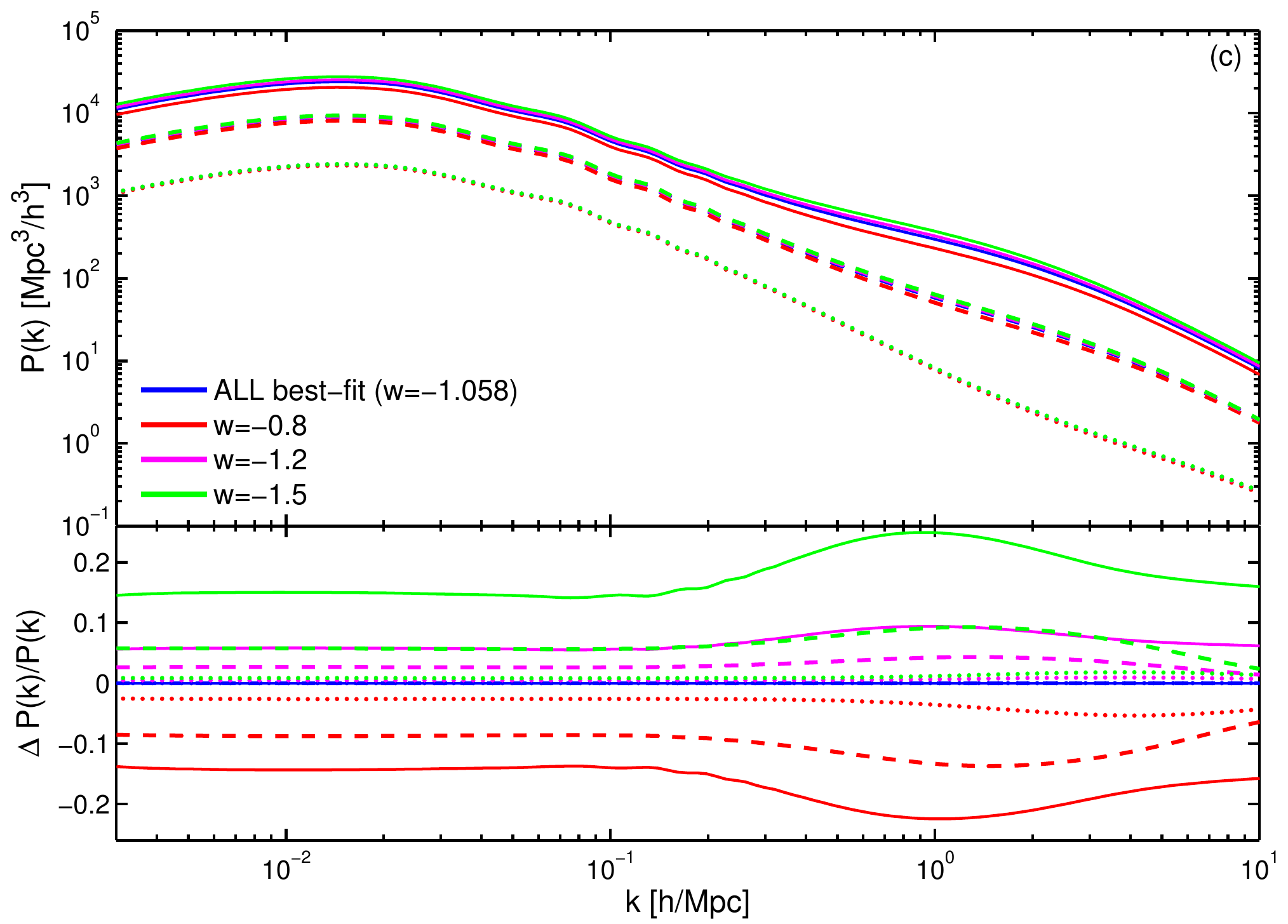}
    \caption{Matter power spectra and their fractional differences. In each panel, the sets of 
solid, dashed and dotted lines represent power spectra at redshift $z=0,1,3$, respectively. 
The blue curves in each panel are calculated from the best-fit parameter values when all cosmological 
probes are considered, with all three extended parameters ($\mnu,~\neff,~w$) kept free 
(see Table~\ref{table:w_mnuNeff}). Compared with the blue set of lines, other sets are computed 
with only one given parameter altered to the values shown within each panel.}
    \label{fig:Pmatt}
\end{figure}

We first display the signatures of massive neutrinos on matter power spectra at redshift $z=0,1,3$, in Fig.~\ref{fig:Pmatt}.
From Panel (a), as neutrinos become heavier, the suppression is increasingly 
severe at small scales, with the maximum suppression in the non-linear regime 
represented roughly by $\Delta P/P\sim-9.8\(\Onu/\Om\)$ \citep{2008JCAP...08..020B}. 
The redshift-dependence of the maximum suppression is 
analytically captured by the HALOFIT--$\nu$ mapping algorithm \citep{2012MNRAS.420.2551B}. 
From a crude glance at Panel (b), as more extra light degrees of freedom are introduced, 
matter power at small scales are more damped as well. Under more cautious inspection, 
the acoustic peaks are meanwhile shifted to smaller scales.
We attempt to interpret these results in what follows.

With a fixed total matter density $\Om h^2$, the increase of $\neff$ will postpone 
matter-radiation equality (see Eq.~\ref{eq:zeq}). Since at subhorizon scales, 
the matter perturbations grow more efficiently during the matter-dominated epoch
rather than during the radiation-dominated epoch, it is reasonable to deduce that the matter power spectrum is 
more severely damped at small scales relatively to large scales, which is one of the most 
pronounced features displayed in Fig.~\ref{fig:Pmatt}(b).  
For CMB, this effect is opposite; the height of the first acoustic peak is increased,
mainly through the early integrated Sachs-Wolfe effect \citep{2008PhRvD..78h3526I}.
On the contrary, for the third and higher order peaks, the amplitudes are more damped due to 
the free-streaming of ultrarelativistic neutrinos \citep{2004PhRvD..69h3002B}.
The position of the $n$th acoustic peak can be estimated by $l_n\sim n\pi/\thA$, 
with the acoustic scale parameter defined as (also see Table~\ref{table:priors}),
\begin{equation}
\thA=\frac{s(z_{\rm dec})}{r(z_{\rm dec})},
\end{equation}
where $r$ is given by Eq.~\ref{eq:rchi} and $z_{\rm dec}\sim1090$ signals the epoch of 
photon-baryon decoupling \citep{2006PhRvD..74l3507T}. Furthermore, $z_{\rm dec}$ is 
insensitive to the presence of massive neutrinos and $r(z_{\rm dec})$ almost remains unaltered 
for different values of $\neff$ \citep{2005PhRvD..71d3001I}. Usually, 
the sound horizon is calculated via \citep{1998ApJ...496..605E},
\begin{equation}
s(z)= \int^{1/(1+z)}_0 c_s(a)\frac{\ud a}{{a}^2H(a)},
\end{equation}
with the sound speed for the photon-baryon fluid $c_s^2=c^2/\[3(1+R)\]$ and $R=\(3\Ob/4\Omega_{\gamma}\)a$.
Since the massive neutrinos we are interested in maintain their relativistic speed before the recombination epoch, 
it is reasonable to conjecture that every neutrino species can be counted as a part of the radiation during that period.
By increasing (decreasing) the value of $\neff$, equivalently $H(a)$ is changed according to Eq.~\ref{eq:expan},
and the sound horizon at the time of decoupling, which dictates the position of the acoustic peaks, 
is thereby shrinked (enlarged). As a consequence, the acoustic peaks are shifted to smaller (larger) scales, 
corresponding to larger (smaller) wavenumbers of the matter power spectrum, particularly shown in Fig.~\ref{fig:Pmatt}(b).
This feature, which is unique to the variation of $\neff$, might provide a smoking gun signature for the existence of 
extra light degrees of freedom.  

Unlike the impacts from ($\mnu,~\neff$), the effect introduced by varying $w$ is a global change of magnitude 
for the matter power spectrum shown in Fig.~\ref{fig:Pmatt}(c).
As inferred from Eq.~\ref{eq:expan}, a larger value for $w$ enables the dark energy
to present itself with a non-vanishing amount in earlier stage of the expansion history. It thereby explains
why the dotted curves in Fig.~\ref{fig:Pmatt}(c), corresponding to matter power spectra at redshift $z=3$, 
are almost immune against the decrease of $w$ deviating from $-1$. The change of $w$ directly modifies 
the expansion history, it is therefore reasonable that the matter power at every scale is simultaneously enhanced/reduced. In addition, as shown, the redshift-dependence of maximum suppresion
is still captured by the fitting algorithm of HALOFIT--$\nu$.
Although the original HALOFIT mapping is only calibrated for the $\Lambda$CDM cosmological model, 
through appropriate modifications of the growth function \citep{2009PhRvD..80b3003J} or a specific fitting prescription
developed by Ref.~\citep{2006MNRAS.366..547M}, one can reliably quantify the non-linear effect of varying $w$,
accurate to a few percent.

\begin{figure}
    \centering
    \includegraphics[width=.6\textwidth]{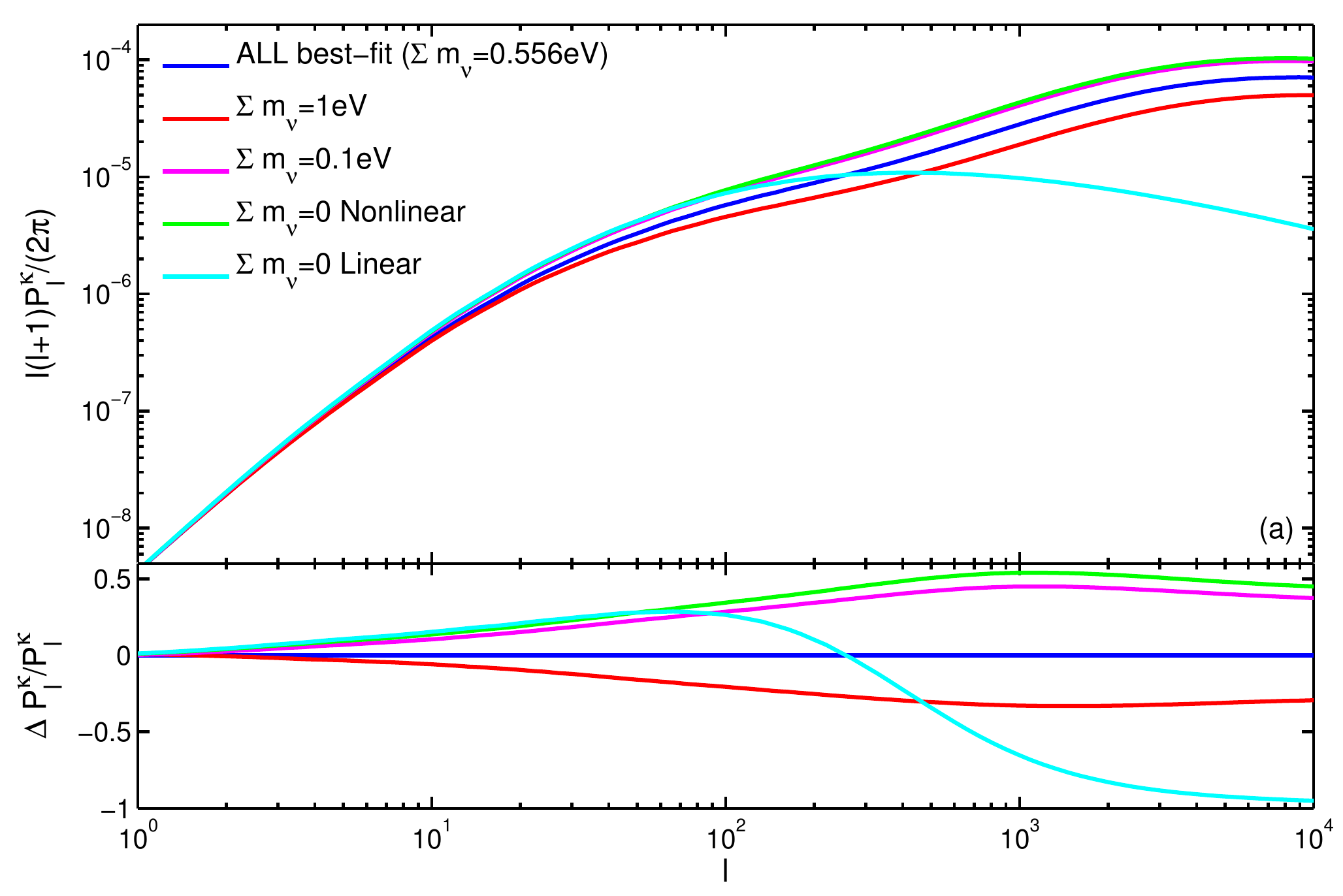}
    \includegraphics[width=.6\textwidth]{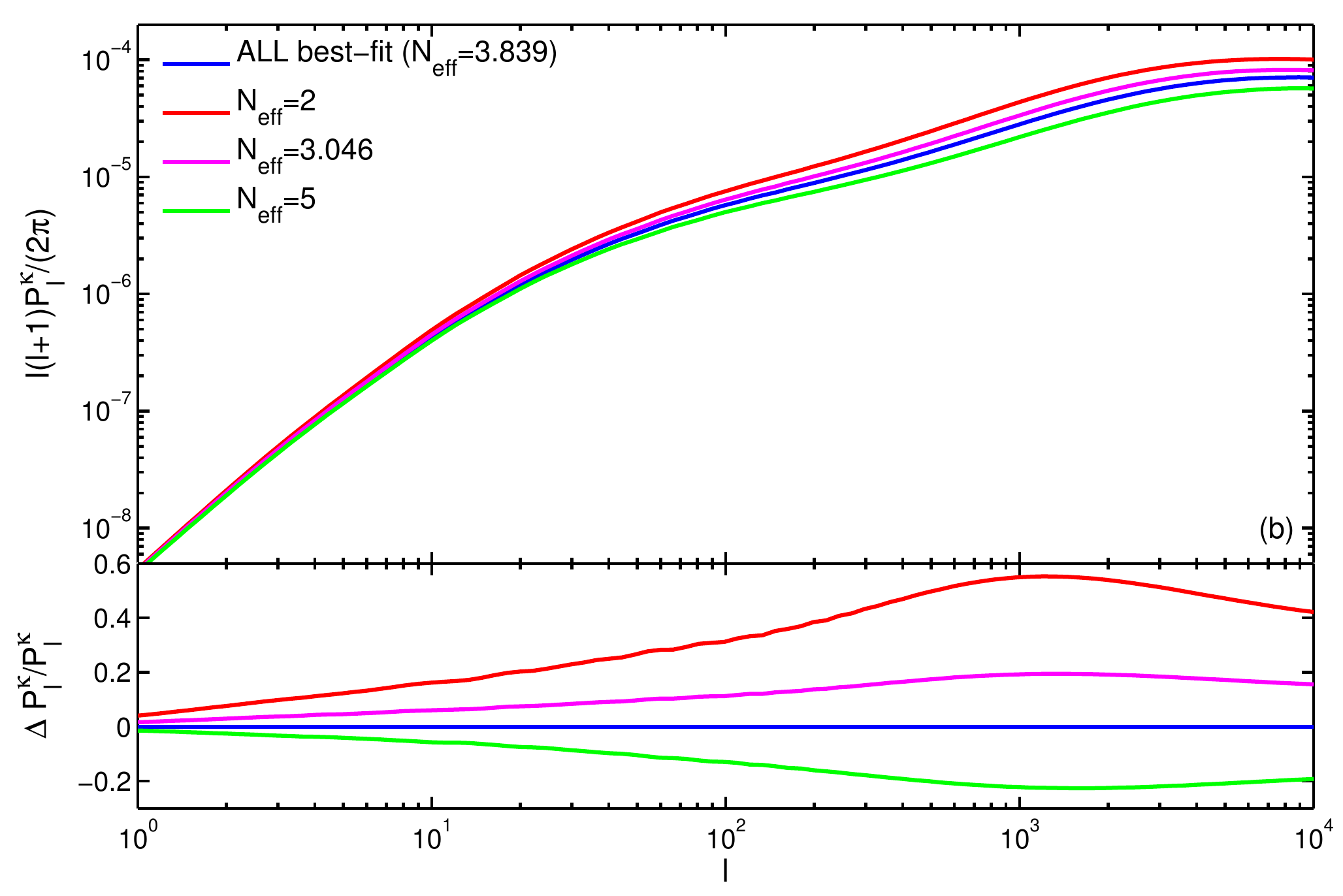}
    \includegraphics[width=.6\textwidth]{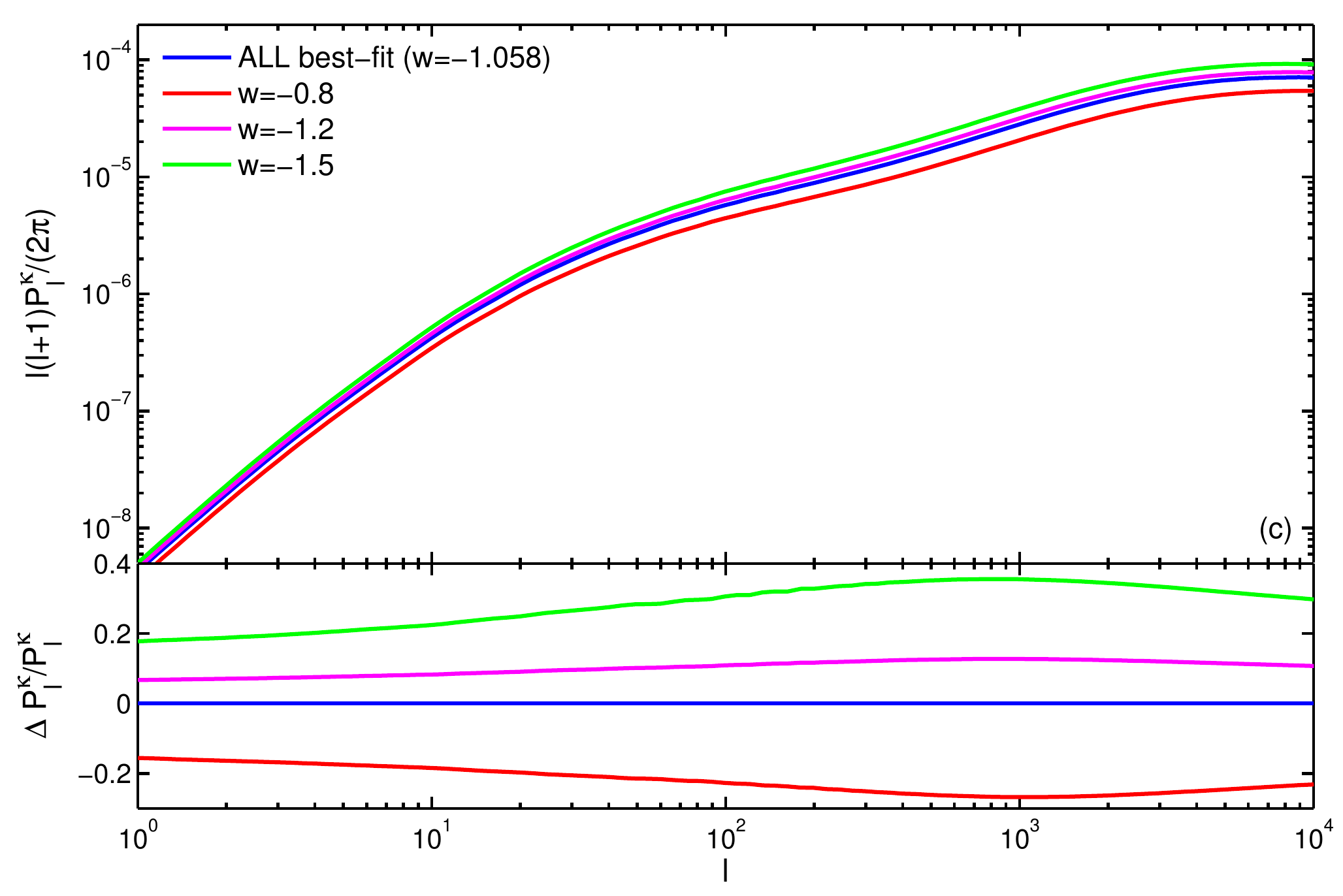}
    \caption{Convergence power spectra and their fractional differences. As in Fig.~\ref{fig:Pmatt},
the blue curve corresponds to the best-fit parameter values given all cosmological data with all of 
($\mnu,~\neff,~w$) free (see Table~\ref{table:w_mnuNeff}). 
Other lines are plotted with only one given parameter changing its value
in order to show the effects of the three extended parameters.}
    \label{fig:Pkappa}
\end{figure}

Fig.~\ref{fig:Pkappa} presents the WL convergence power spectra as calculated from Eq.~\ref{eq:Pkappa}, with each panel
corresponding to the panels of Fig.~\ref{fig:Pmatt}. 
No wonder the convergence power spectrum is also reduced by the increase of $\mnu$ and $\neff$, or the decrease of $w$,
since it is given by the integraion of matter power spectra along the line of sight. 
After integration, we have lost the unique signature for $\neff$.
Fortunately, the impact from $w$ still holds the different behavior of a global change, which can be used to
differentiate from the effects introduced by $\mnu$ and $\neff$. 
The maximum suppression on the convergence power is at $l\sim10^3$, insensitive to the variations of the three parameters.
Besides, we also plot the result predicted by the linear perturbation theory in Fig.~\ref{eq:Pkappa}(a).
The primary discrepancy takes place below the scale corresponding to $l>10^2$, and it thereby illustrates the importance of an accurate knowledge of the full non-linear matter power spectrum, 
for future WL surveys to achieve their full potential \citep{2005APh....23..369H}.   

Eventually, the numerical results of the aperture-mass variance given by Eq.~\ref{eq:Map} 
are demonstrated in Fig.~\ref{fig:Map}, together with the actual data of CFHTLS-T0003.
From Panel (a), we see that the current detection of WL signals mainly lies within the semi- 
to non-linear regime, where the non-linear modeling of matter clustering is vital for a precise interpretation of the observational data.
As repeatedly mentioned, the HALOFIT-$\nu$ algorithm has great advantages itself, yet the effects of baryons are not covered. 
The hot baryons harbored within intracluster medium have been shown to affect the WL shear 
power spectrum up to ten percent at $k\sim10\rm~h/Mpc$ \citep{2004ApJ...616L..75Z}. 
If sufficiently strong AGN feedback is included, ten percent is already reached at $k\sim1\rm~h/Mpc$ \cite{2011MNRAS.415.3649V}.
Using hydrodynamic simulations with consideration of baryon physics, Ref.~\cite{2011MNRAS.417.2020S} discovered that 
as much as 40 percent bias in the constraint on dark energy EOS can be introduced by ignoring the effects of baryons. 
However, most of the baryon contribution is at smaller scales than currently observed. 
Besides, finite-mass neutrinos exert quite different impacts on matter clustering compared with the baryons. 
Thus using current optimal mapping algorithm, HALOFIT-$\nu$, 
will suffice for the constraints on $\mnu$. 
In the precision analysis of the WL, not only are the theoretical prescriptions crucial, 
but the data quality is also influential. As shown in Fig.~\ref{fig:Map}, a small bump 
is present around the semi-nonlinear scales in the CFHTLS data. Although $\map$ is a more localized filter and thus less prone to large scale systematics or E-/B-mode mixing, this feature is still mainly caused by some thorny systematic issues,
e.g. shape measurements, intrinsic alignments, PSF anisotropies, catastrophic photo-$z$ error.
For a thorough discussion on WL systematics, observables and their interplay, we refer the reader to Ref.~\citep{2006MNRAS.366..101H}.
Anyhow, the community is looking forward to better settlement of these systematics, in the final release of CFHTLS lensing data in the near future.

\section{Conclusion and Discussion}\label{sec:conclu}
In this paper, we present a comprehensive analysis of the constraints on the sum of neutrino masses, 
the effective number of neutrino species, and the constant EOS parameter of dark energy.
Via employing the most recent observational data of CMB+WL+BAO+OHD+SNIa,
we have pinned down the total neutrino mass to $\mnu<0.476$ at 95\% CL,
whereas the constraint reads $\mnu=0.421^{+0.186}_{-0.219}$ (68\% CL) when $\neff$ is simultaneously set loose,
and even gives $\mnu=0.556^{+0.231}_{-0.288}$ (68\% CL) if $w$ is also included as a free parameter.
Given that we only consider the constant EOS for the dark energy in present work, 
the results including evolving dark energy component will provide even less stringent constraints.
This highlights our concern that in constraining neutrino masses using cosmological probes, 
one needs to take into consideration the correlations between ($\mnu,~\neff,~w$) seriously.
For the effective number parameter, we obtained the best constraints as $\neff=3.217\pm0.367$ (with fixed $\mnu$ and $w$),
$\neff=3.740\pm0.446$ (with freed $\mnu$ and fixed $w$), $\neff=3.454\pm0.386$ (with fixed $\mnu$ and freed $w$),
and $\neff=3.839\pm0.452$ (with both $\mnu$ and $w$ freed).
For the most generalized parameter scenario, the standard model of $\neff=3.046$ is compatible at $1.75\sigma$ CL.
If statistical uncertainties were dominating, which is probably true for the BAO and OHD, 
then we can rule out the standard three flavor neutrinos model at $5\sigma$ CL,
when the future datasets over eight times as large as current samples are obtained.
But for some probes like SNIa, systematics are already dominating, 
so the expected improvements are not easily estimated. 
Most probably in the future, equivalent or improved constraints will be obtained 
with BAO angular diameter distances and Hubble parameter measurements.
The constraints on the EOS parameter all along favor the cosmological constant model,     
regardless of whether or not varying $\mnu$ and/or $\neff$. 
That is mainly due to the effectiveness of the SNIa data.
Moreover, the OHD still provide helpful improvements on dark energy constraints. 
We also pay particular attention to parameter degeneracies.
The strong correlation between $\neff$ and $H_0$ (also $\neff$ and $\Och$) can be explained by the rather
certain epoch for matter-radiation equality.
The strong anticorrelation found in ($\mnu$ vs. $\sigma_8$) results from the suppression of matter 
perturbation growth ascribed to the presence of finite-mass cosmic neutrinos.
As shown in Sect.~\ref{sec:constrain}, due to the different proportionality trends that $\mnu$ and $\neff$ hold with respect to $\sigma_8$, as well as the advantage of WL in measuring $\sigma_8$, the technique of WL does show great potential in discriminating between the effects of $\mnu$ and $\neff$ for future large surveys.

Another key aspect in our work is the investigation on the constraining capabilities of different cosmological probes.
The WL observation at current stage does contribute to some improvement if the dark energy parameter is fixed.
However when $w$ is free to vary, adding the WL data does not result in much better constraints on $\mnu$. 
Perhaps the WL tomography and its cross-correlation with other tomographic probes will help 
improve this situation \citep{2006JCAP...06..025H, 2012PhRvD..86b3526J}.
On the comparison between the results from the inclusions of OHD and SNIa, 
it is found that for the constraints on $\mnu$, 
when dark energy EOS is fixed ($w=-1$), the result given by OHD is similar
to that by SNIa luminous distances or BAO angular diameter distances, with OHD slightly better. 
If $\neff$ is set free, OHD distinctly prevail over SNIa.
Nevertheless, as long as $w$ is freed, the SNIa data confine $\mnu$ more strictly than the OHD do.
Concerning the constraints on the Hubble constant, 
when only $\mnu$ is set loose, the SNIa data provide similar constraints as the OHD do.
Once $\neff$ is varying, the OHD perform much better than the SNIa,
due to the overwhelming power of the OHD in constraining $\neff$.
Additionally, the SNIa and the OHD provide almost the same constraints on $\Om$.
On the degeneracy between $\neff$ and $w$, the inclusions of OHD and SNIa
lead to nearly orthogonal degeneracy directions, as explained by their 
specific powers in constraining $\neff$ and $w$, respectively. As a consequence, their union will
facilitate the breaking of this parameter degeneracy so well that our global constraints on all cosmological parameters will benefit. 

The tantalizing indication of a non-zero total neutrino mass is really interesting 
(see Sec.~\ref{subsec:l_mN} and Sec.~\ref{subsec:w_mN}). We argue this is the result of the 
combined effect from all cosmological probes in our study. 
First of all, due to the large parameter space and the fact that some parameters, e.g. $\Obh$ and 
$\tau$, are badly constrained by other probes, the CMB data are indispensable, yet not
enough even when combined with the $H_0$ prior. The WL technique provides plenty of
information on mass distribution and matter clustering, which are sensitive to the neutrino total
mass, however its use is limited by measurement errors as well as the inclusion of variable $w$.
Although not directly related to neutrino masses, the late-time probes, i.e. BAO, SNIa and OHD, 
are particularly helpful in disentangling some parameter degeneracies. 
Thereby combining all the observations together, i.e. CMB+WL+BAO+OHD+SNIa(+HST), we can  
constrain $\mnu$ very precisely and even impose a 68\% CL lower limit.  
As also pointed out recently by Ref.~\cite{2012arXiv1209.1043H}, neither future CMB experiment 
or weak lensing shear measurement or galaxy tomographic survey alone can reach $\sim5\sigma$ 
detection of a non-zero sum of neutrino masses; 
it is their joint analysis that effectively breaks parameter degeneracies. 

The different impacts that dark energy, cosmic neutrino total mass and effective number exert onto 
matter clustering and structure formation are also discussed. 
Regarding the effects on low-redshift matter power spectra, the increase of $w$ tends to 
reduce the global magnitude of matter power at all scales, due to direct modification of the expansion law.
On the contrary, the change in $\mnu$ and/or $\neff$ only affects the power spectrum at scales $k>k_{\rm NR}$.
The increase of $\mnu$ damps more severely the non-linear matter power due to the free-streaming effect, 
while the increment on $\neff$ causes delay to the matter-radiation equality epoch, and thus
suppresses the growth of matter perturbation at small scales. The suppresion caused by $\neff$ is 
nearly redshift-independent for $k<1\rm~h/Mpc$. 

In addition, we note an unique feature of $\neff$: 
its variation is capable of shifting acoustic peaks in late-time matter power spectra.
Accordingly, the WL convergence power spectrum is altered, and the aperture-mass variance affected.
It is also discussed that the precision analysis of WL demands the accurate knowledge of the full
non-linear matter power spectrum and the next stage cosmological simulations had better take gas 
dynamics into account. Albeit the current data of CFHTLS still suffer from systematics, high hope
and great endeavor have been invested in this program, which will finally promote our knowledge on 
the formation and evolution of cosmic structure through WL.

Recently, Ref.~\cite{2012arXiv1208.4354J} tested the model of eV mass scale sterile neutrinos, 
employing two different SNIa datasets given by different light curve fitters, 
i.e. Union2 with SALT2, and SDSS-II with MLCS2k2. They found that such sterile neutrino model 
is strongly supported by the SNIa sample of SDSS-II with MLCS2k2 ($\sim$ 55 times more probable 
than the null model), whereas disfavored by the Union2 SNIa data with SALT2, as well as other 
cosmological probes. This is not entirely unexpected since several studies have already 
shown that MLCS2k2 seems to support more exotic physics rather than the $\Lambda$CDM model, 
whereas the very same data standardized with SALT2 are fairly consistent with cosmological 
constant as well as other types of cosmological observations \cite{2009ApJ...703.1374S, 2009JCAP...11..029B, 2012ApJ...744..176L}. 
We note that in judging the viability of Ref.~\cite{2012arXiv1208.4354J}'s sterile
neutrino model, the different fitters do play a role, but also the different SNIa data
samples and the estimates of the systematic errors associated with the samples. Therefore to conclude that the light curve fitter is the key factor of vindicating such sterile neutrino model or not,
more work should be dedicated to the re-fitting of the same SNIa dataset, especially Union2.1, using different methods.

Cosmological probes we have discussed are mainly sensitive to the sum of neutrino masses, 
yet almost blind to mass splitting scenarios, if any, for different neutrino eigenstates. 
Fortunately, cosmological observations can reliably rule out total degeneracy between 
the eigenstates if $\mnu$ is constrained \textit{robustly} to be less than 0.22 eV, and furthermore nail down the Normal Hierarchy scenario as the 
true mass splitting when $\mnu<0.1\rm~eV$ is \textit{conservatively} derived \citep[cf. Ref.][Figure~1]{2011JCAP...03..030C}.

\begin{figure}
    \centering
    \includegraphics[width=.6\textwidth]{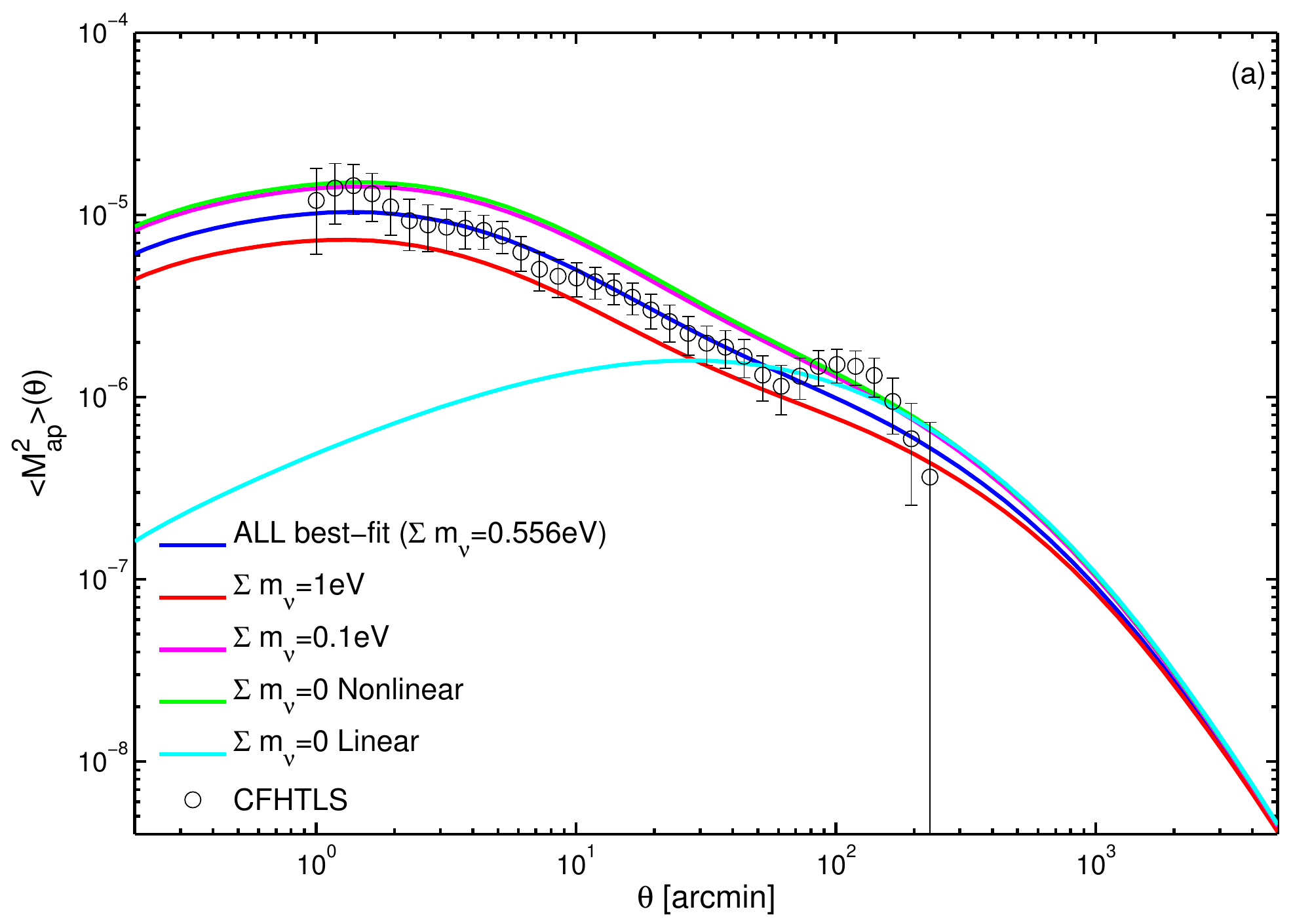}
    \includegraphics[width=.6\textwidth]{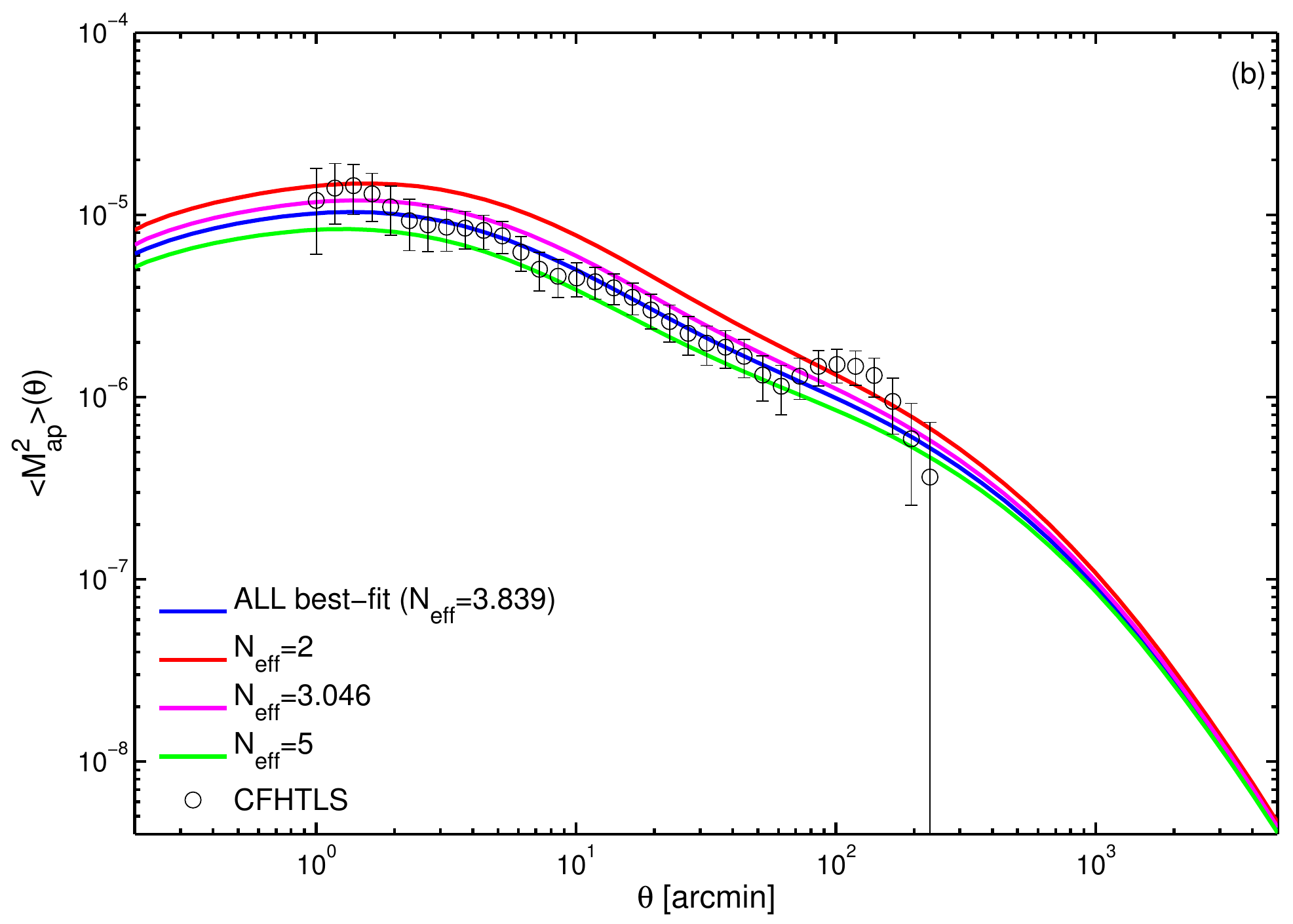}
    \includegraphics[width=.6\textwidth]{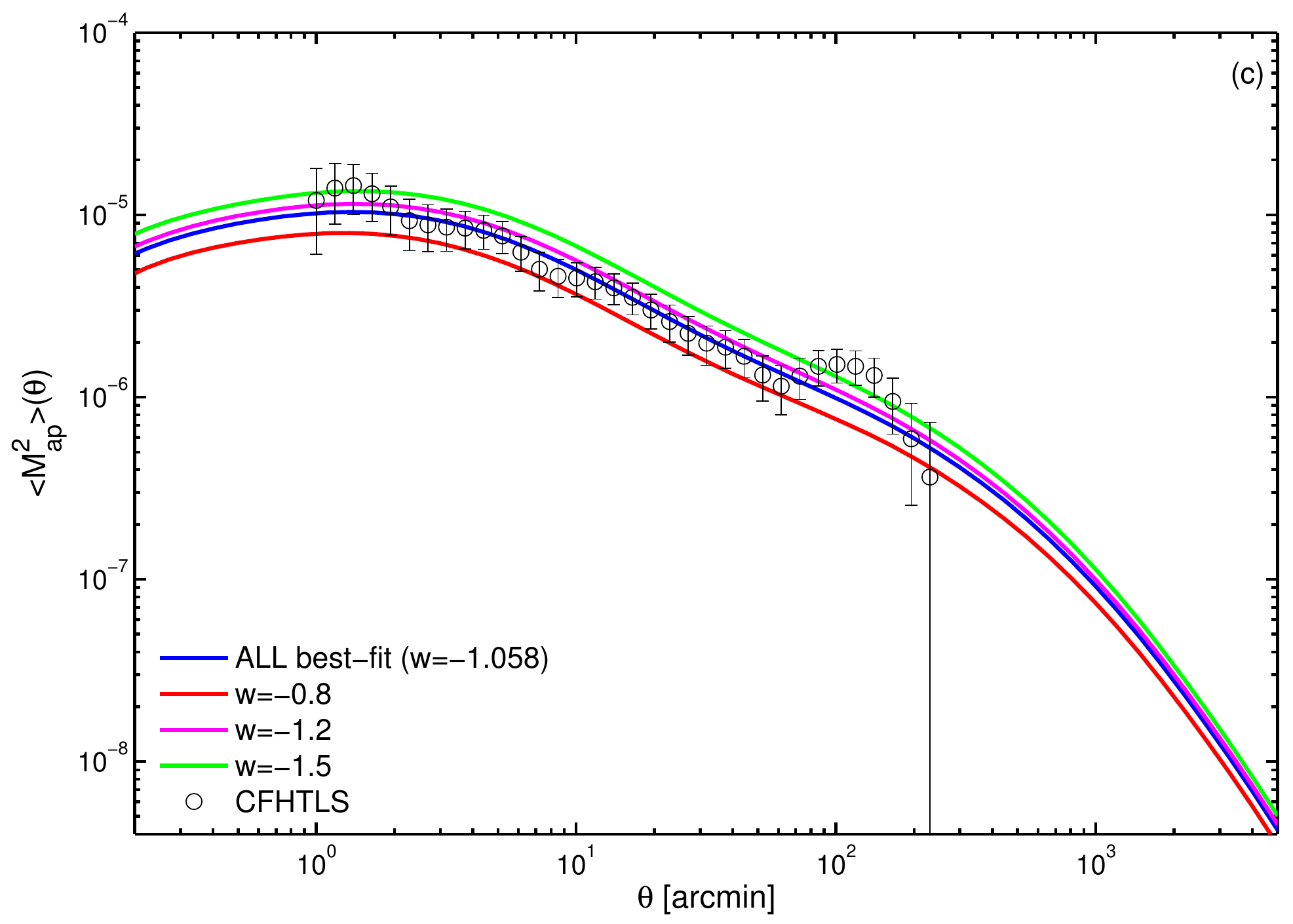}
    \caption{Aperture-mass variance from observation (CFHTLS) and from theoretical calculations. 
As in Fig.~\ref{fig:Pmatt}, the blue curve is given by the best-fit parameter values 
against the full combination with all of ($\mnu,~\neff,~w$) free (see Table~\ref{table:w_mnuNeff}).
Other lines are drawn with only one given parameter altered.}
    \label{fig:Map}
\end{figure}

However, as shown in Fig.~\ref{fig:Map}, these small values for $\mnu$ are seemingly disfavored by 
the $\map$ data alone, as the constraint on $\neff$ tends (slightly) to prefer additional light 
degrees of freedom. In practice, there have been some arguments on discriminating the neutrino mass 
hierachy within the framework of Fisher formalism \citep{2006PhRvD..73l3501S, 2009PhRvD..80l3509D, 2010JCAP...05..035J, 2012MNRAS.tmp.3504H}. 
Ref.~\citep{2012ApJ...752L..31W} conducted N-body simulations to study seperately the effects from 
different mass hierarchies and obtained a $\sim0.5\%$ difference. This small, yet non-trivial 
result underlines its measurable potential from upcoming space surveys (e.g. Euclid), in 
combination with ground-based experiments. In our present analyses, $\neff$ is treated as the 
effective number of neutrinos. We do realize that except members of the neutrino family, 
other more exotic candidates, e.g. axions, gravitinos, or even primordial 
magnetic fields, can equally fill for this extra relativistic component.
Besides, early dark energy has been shown to be able to reconcile 
the standard model of three families of neutrinos with various cosmological observations 
\cite{2012arXiv1202.0005J}.
Nonetheless, as large scientific programs are already providing new results, 
we are confident that the truth of dark radiation will soon be unveiled.

\acknowledgments
CFHTLS is based on observations obtained with {\sc MegaPrime/MegaCam}, a
joint project of CFHT and CEA/DAPNIA, at the Canada-France-Hawaii
Telescope (CFHT) which is operated by the National Research Council
(NRC) of Canada, the Institut National des Sciences de l'Univers of
the Centre National de la Recherche Scientifique (CNRS) of France,
and the University of Hawaii.  This work is based in part on data
products produced at {\sc Terapix} and the Canadian Astronomy Data
Centre as part of the Canada-France-Hawaii Telescope Legacy Survey,
a collaborative project of NRC and CNRS.
We acknowledge the use of the Legacy Archive for Microwave Background
Data Analysis (LAMBDA). Support for LAMBDA is provided by the NASA
Office of Space Science. The use of the user-friendly and
freely available package CosmoMC and CAMB is also warmly acknowledged. 
The numerical calculations in this paper have been done on the IBM 
Blade cluster system in the High Performance Computing Center (HPCC) of 
Nanjing University.

The useful suggestions from the anonymous referee are greatly appreciated.
We are deeply grateful to the helpful conversations with Hao Wang, Qiao Wang, Cong Ma,
Lei Sun, You-Hua Xu, Shuo Yuan, Jun-Qing Xia, Gong-Bo Zhao, Hu Zhan, Hong Li, Liang Cao, 
Shahab Joudaki, Martin Kilbinger, Philip Marshall, Richard Shaw, and Joop Schaye.
We thank Antony Lewis for technical support on cosmocoffee \footnote{http://cosmocoffee.info}.
This work was supported by the National Science Foundation of China (grant No. 11173006), 
the Ministry of Science and Technology National Basic Science program (project 973) 
under grant No. 2012CB821804, the Fundamental Research Funds for the Central Universities, 
the National Natural Science Foundation of China (Grant No. 11033002)
and the National Basic Research Program of China (973 Program, Grant No. 2009CB824800).

\bibliographystyle{JHEP}
\bibliography{mvn}







\begin{table}
\begin{center}
\caption{Key parameters and their prior information.}
\begin{tabular}{llc|c}
\hline\hline
Set & Parameter & Symbol & Prior\\
\hline
Vanilla & Physical baryon density & $\Obh$ & $[0.005,0.1]$\\
& Physical dark matter density & $\Och$ & $[0.01,0.99]$\\
& $100~\times~$Angular size of sound horizon & $100\times\thA$ & $[0.5,10]$\\
& Reionization optical depth & $\tau$ & $[0.01,0.8]$\\
& Scalar spectral index & $\ns$ & $[0.5,1.5]$\\
& Scalar spectral amplitude & $\ln{(10^{10} \As)}$ & $[2.7,4]$\\
\hline
Extended & Sum of neutrino masses & \mnu~[eV] &  $[0,10]$ \\
& Effective number of neutrinos & \neff &  $[0,10]$ \\
& Constant dark energy EOS & $w$ & $[-2,0]$ \\
\hline
Derived & Total matter density  &  $\Om$  &  --- \\
& Power spectrum normalization & $\sigma_8$  & --- \\
& Hubble constant  & $H_0$  &  --- \\ 
\hline\hline
\end{tabular}
\begin{tablenotes}
\item The ``Vanilla'' set of parameters is the basic set that is always 
fed into our MCMC code with uniform prior distribution, as shown within the table. 
What we are exceedingly interested in is the ``Extended'' set of parameters, whose flat prior distributions 
are also given. By fixating some members of this set while varying the rest, 
it is possible to investigate individually and collectively the constraints they receive 
from observational data. And meanwhile we are able to learn some knowledge about 
how parameter degeneracies affect the constraints.
As shown in this table, $\Om,~\sigma_8$ and $H_0$ are always derived.
Additionaly, a flat prior ($[0,2]$) is assigned to $A_{\textrm{sz}}$, i.e.
the SZ template amplitude, which is always marginalized as a nuisance parameter.
\end{tablenotes}
\label{table:priors}
\end{center}
\end{table}

\clearpage
\begin{table*}\footnotesize
\begin{center}
\caption{Constraints on Parameter Scenario of ``Vanilla+$\mnu$'' Using Multiple Combinations of Cosmological Probes.}
\begin{tabular}{lcccccccccccccc|}
\hline \hline
&  & CMB & CMB+WL & CMB+BAO+OHD & CMB+BAO+SNIa & CMB+WL+BAO+OHD+SNIa \\
\hline \hline
Vanilla & $100\Obh$	& $2.242 \pm 0.055$ & $2.241 \pm 0.051$ & $2.242 \pm 0.051$ & $2.236 \pm 0.052$ & $2.229 \pm 0.050$ \cr
& $100\Och$		& $11.04 \pm 0.49 $ & $11.00 \pm 0.35 $ & $11.33 \pm 0.32 $ & $11.33 \pm 0.31 $ & $11.19 \pm 0.24 $  \cr
& $10^4\thA$		& $103.91\pm 0.25 $ & $103.90\pm 0.25 $ & $103.94\pm 0.25 $ & $103.91\pm 0.26 $ & $103.88\pm 0.23 $  \cr
& $\tau$ 		& $0.090 \pm 0.015$ & $0.089 \pm 0.015$ & $0.088 \pm 0.014$ & $0.087 \pm 0.015$ & $0.087 \pm 0.014$  \cr
& $n_s$ 		& $0.970 \pm 0.013$ & $0.969 \pm 0.013$ & $0.968 \pm 0.013$ & $0.967 \pm 0.012$ & $0.964 \pm 0.011$ \cr
&$\ln{(10^{10} \As)}$   & $3.073 \pm 0.036$ & $3.070 \pm 0.031$ & $3.081 \pm 0.034$ & $3.076 \pm 0.035$ & $3.068 \pm 0.030$  \cr
\hline
Extended & $\mnu$ [eV]  & $<0.524$	    & $<0.496$ 		& $<0.486$	    & $<0.518$ 		& $<0.476$     \cr
\hline
Derived & $\Om$ 	& $0.272 \pm 0.028$ & $0.270 \pm 0.023$ & $0.286 \pm 0.015$ & $0.288 \pm 0.015$ & $0.283 \pm 0.013$ \cr 
& $\sigma_8$ 		& $0.755 \pm 0.046$ & $0.751 \pm 0.035$ & $0.766 \pm 0.048$ & $0.758 \pm 0.049$ & $0.743 \pm 0.035$ \cr
& $H_0/100$ 		& $0.700 \pm 0.024$ & $0.702 \pm 0.022$ & $0.689 \pm 0.012$ & $0.687 \pm 0.013$ & $0.689 \pm 0.012$ \cr
\hline
\hline
\end{tabular}
\begin{tablenotes}
\item Concerning the reported results of all parameters except \mnu, 
the best-fit values refer to the means calculated 
from one-dimensional marginalized posterior probability distributions, 
as plotted in Fig.~\ref{fig:lam_mnu} for some parameters.
The reported errors originate from the symmetric 68\% confidence intervals
with respect to the corresponding means. For $\mnu$, the 95\% upper limit is given. 
Here $\neff$ and $w$ are fixed to their standard values ($\neff=3.046,~w=-1$).
\end{tablenotes}
\label{table:lam_mnu}
\end{center}
\end{table*}

\begin{table*}\footnotesize
\begin{center}
\caption{Constraints on Parameter Scenario of ``Vanilla+$\neff$'' Using Multiple Combinations of Cosmological Probes.}
\begin{tabular}{lcccccccccccccc|}
\hline \hline
&  & CMB & CMB+WL & CMB+BAO+OHD & CMB+BAO+SNIa & CMB+WL+BAO+OHD+SNIa \\
\hline \hline
Vanilla & $100\Obh$	& $2.215 \pm 0.057$ & $2.228 \pm 0.057$ & $2.214 \pm 0.054$ & $2.213 \pm 0.052$ & $2.204 \pm 0.052$ \cr
& $100\Och$		& $13.54 \pm 1.85 $ & $11.15 \pm 0.76 $ & $12.56 \pm 0.83 $ & $14.50 \pm 1.50 $ & $11.39 \pm 0.57 $  \cr
& $10^4\thA$		& $103.42\pm 0.38 $ & $103.74\pm 0.37 $ & $103.57\pm 0.32 $ & $103.30\pm 0.35 $ & $103.64\pm 0.33 $  \cr
& $\tau$ 		& $0.086 \pm 0.015$ & $0.089 \pm 0.015$ & $0.084 \pm 0.014$ & $0.084 \pm 0.014$ & $0.081 \pm 0.013$  \cr
& $n_s$ 		& $0.978 \pm 0.014$ & $0.971 \pm 0.014$ & $0.969 \pm 0.012$ & $0.982 \pm 0.014$ & $0.961 \pm 0.011$ \cr
&$\ln{(10^{10} \As)}$   & $3.126 \pm 0.048$ & $3.074 \pm 0.032$ & $3.105 \pm 0.035$ & $3.139 \pm 0.040$ & $3.064 \pm 0.029$  \cr
\hline
Extended & $\neff$ 	& $4.340 \pm 0.817$ & $3.334 \pm 0.496$ & $3.722 \pm 0.418$ & $4.827 \pm 0.843$ & $3.271 \pm 0.367$ \cr
\hline
Derived & $\Om$ 	& $0.283 \pm 0.031$ & $0.250 \pm 0.013$ & $0.287 \pm 0.015$ & $0.289 \pm 0.015$ & $0.270 \pm 0.009$ \cr 
& $\sigma_8$ 		& $0.870 \pm 0.053$ & $0.799 \pm 0.023$ & $0.848 \pm 0.029$ & $0.892 \pm 0.041$ & $0.804 \pm 0.018$ \cr
& $H_0/100$ 		& $0.746 \pm 0.026$ & $0.732 \pm 0.024$ & $0.717 \pm 0.016$ & $0.760 \pm 0.032$ & $0.710 \pm 0.016$ \cr
\hline
\hline
\end{tabular}
\begin{tablenotes}
\item Same as Table~\ref{table:lam_mnu}, except that $\neff$ is free to vary,
with massless neutrinos and $w=-1$. We present the results of $\neff$ following the usual convention
as introduced in Table~\ref{table:lam_mnu}. 
\end{tablenotes}
\label{table:lam_Neff}
\end{center}
\end{table*}

\begin{table*}\footnotesize
\begin{center}
\caption{Constraints on Parameter Scenario of ``Vanilla+$\mnu$+$\neff$'' Using Multiple Combinations of Cosmological Probes.}
\begin{tabular}{lcccccccccccccc|}
\hline \hline
&  & CMB & CMB+WL & CMB+BAO+OHD & CMB+BAO+SNIa & CMB+WL+BAO+OHD+SNIa \\
\hline \hline
Vanilla & $100\Obh$	& $2.186 \pm 0.058$ & $2.204 \pm 0.059$ & $2.214 \pm 0.055$ & $2.212 \pm 0.054$ & $2.213 \pm 0.055$ \cr
& $100\Och$		& $16.75 \pm 2.95 $ & $13.32 \pm 1.89 $ & $12.68 \pm 0.86 $ & $14.35 \pm 1.42 $ & $12.34 \pm 0.80 $  \cr
& $10^4\thA$		& $103.11\pm 0.34 $ & $103.43\pm 0.36 $ & $103.56\pm 0.32 $ & $103.35\pm 0.35 $ & $103.58\pm 0.32 $  \cr
& $\tau$ 		& $0.084 \pm 0.014$ & $0.086 \pm 0.014$ & $0.088 \pm 0.015$ & $0.087 \pm 0.014$ & $0.087 \pm 0.014$  \cr
& $n_s$ 		& $0.985 \pm 0.014$ & $0.976 \pm 0.014$ & $0.974 \pm 0.013$ & $0.987 \pm 0.015$ & $0.972 \pm 0.012$ \cr
&$\ln{(10^{10} \As)}$   & $3.157 \pm 0.050$ & $3.097 \pm 0.038$ & $3.101 \pm 0.037$ & $3.132 \pm 0.041$ & $3.086 \pm 0.033$  \cr
\hline
Extended & $\mnu$ [eV]  & $<1.515$ 	    & $<1.393$ 		& $<0.758$  	    & $<0.775$ 	& $0.421_{-0.219}^{+0.186}$  \cr
	 & $\neff$ 	& $5.729 \pm 1.274$ & $4.308 \pm 0.924$ & $3.843 \pm 0.439$ & $4.814 \pm 0.889$ & $3.740 \pm 0.446$ \cr
\hline
Derived & $\Om$ 	& $0.341 \pm 0.051$ & $0.297 \pm 0.040$ & $0.296 \pm 0.018$ & $0.294 \pm 0.015$ & $0.292 \pm 0.016$ \cr 
& $\sigma_8$ 		& $0.806 \pm 0.063$ & $0.729 \pm 0.051$ & $0.762 \pm 0.059$ & $0.807 \pm 0.063$ & $0.731 \pm 0.037$ \cr
& $H_0/100$ 		& $0.746 \pm 0.034$ & $0.724 \pm 0.032$ & $0.709 \pm 0.017$ & $0.749 \pm 0.031$ & $0.707 \pm 0.017$ \cr
\hline
\hline
\end{tabular}
\begin{tablenotes}
\item Same as Table~\ref{table:lam_mnu}, but for both $\mnu$ and $\neff$ kept free, 
with $w=-1$. The conventions for presenting the results of $\mnu$ and $\neff$ are as usual.
Only for the full data combination (CMB+WL+BAO+OHD+SNIa), the result of $\mnu$ is represented by the
best-fit value with asymmetric errors marking the ranges of 68\% confidence interval.
\end{tablenotes}
\label{table:lam_mnuNeff}
\end{center}
\end{table*}

\begin{table*}\footnotesize
\begin{center}
\caption{Constraints on Parameter Scenario of ``Vanilla+$w$+$\mnu$'' Using Multiple Combinations of Cosmological Probes.}
\begin{tabular}{lcccccccccccccc|}
\hline \hline
&  & CMB & CMB+WL & CMB+BAO+OHD & CMB+BAO+SNIa & CMB+WL+BAO+OHD+SNIa \\
\hline \hline
Vanilla & $100\Obh$	& $2.191 \pm 0.059$ & $2.186 \pm 0.062$ & $2.214 \pm 0.054$ & $2.218 \pm 0.052$ & $2.225 \pm 0.054$ \cr
& $100\Och$		& $11.87 \pm 0.70 $ & $11.59 \pm 0.57 $ & $11.82 \pm 0.45 $ & $11.53 \pm 0.43 $ & $11.29 \pm 0.31 $  \cr
& $10^4\thA$		& $103.79\pm 0.26 $ & $103.74\pm 0.26 $ & $103.87\pm 0.25 $ & $103.85\pm 0.26 $ & $103.88\pm 0.24 $  \cr
& $\tau$ 		& $0.085 \pm 0.014$ & $0.086 \pm 0.014$ & $0.085 \pm 0.014$ & $0.087 \pm 0.015$ & $0.086 \pm 0.014$  \cr
& $n_s$ 		& $0.955 \pm 0.015$ & $0.954 \pm 0.015$ & $0.959 \pm 0.013$ & $0.962 \pm 0.013$ & $0.964 \pm 0.012$ \cr
&$\ln{(10^{10} \As)}$   & $3.075 \pm 0.034$ & $3.063 \pm 0.030$ & $3.081 \pm 0.034$ & $3.077 \pm 0.036$ & $3.067 \pm 0.031$  \cr
\hline
Extended & $\mnu$ [eV] 	& $<1.079$	    & $<1.072$ 		& $<0.819$	    & $<0.688$ 		& $<0.627$	    \cr
& $w$ 			& $-1.379\pm 0.247$ & $-1.340\pm 0.237$ & $-1.240\pm 0.182$ & $-1.074\pm 0.088$ & $-1.034\pm 0.080$ \cr
\hline
Derived & $\Om$ 	& $0.255 \pm 0.029$ & $0.250 \pm 0.028$ & $0.268 \pm 0.018$ & $0.286 \pm 0.016$ & $0.283 \pm 0.014$ \cr 
& $\sigma_8$ 		& $0.788 \pm 0.068$ & $0.757 \pm 0.047$ & $0.794 \pm 0.065$ & $0.762 \pm 0.062$ & $0.741 \pm 0.035$ \cr
& $H_0/100$ 		& $0.746 \pm 0.038$ & $0.746 \pm 0.038$ & $0.725 \pm 0.029$ & $0.693 \pm 0.017$ & $0.692 \pm 0.016$ \cr
\hline
\hline
\end{tabular}
\begin{tablenotes}
\item Same as Table~\ref{table:lam_mnu}, but also with $w$ freed, and $\neff$ is still fixed.
The results of $w$ is presented in the conventional manner.
\end{tablenotes}
\label{table:w_mnu}
\end{center}
\end{table*}

\begin{table*}\footnotesize
\begin{center}
\caption{Constraints on Parameter Scenario of ``Vanilla+$w$+$\neff$'' Using Multiple Combinations of Cosmological Probes.}
\begin{tabular}{lcccccccccccccc|}
\hline \hline
&  & CMB & CMB+WL & CMB+BAO+OHD & CMB+BAO+SNIa & CMB+WL+BAO+OHD+SNIa \\
\hline \hline
Vanilla & $100\Obh$	& $2.227 \pm 0.061$ & $2.230 \pm 0.061$ & $2.207 \pm 0.053$ & $2.215 \pm 0.054$ & $2.212 \pm 0.054$ \cr
& $100\Och$		& $17.09 \pm 3.72 $ & $11.02 \pm 1.61 $ & $12.52 \pm 0.84 $ & $14.05 \pm 1.56 $ & $11.50 \pm 0.57 $  \cr
& $10^4\thA$		& $103.16\pm 0.43 $ & $103.82\pm 0.49 $ & $103.60\pm 0.32 $ & $103.36\pm 0.35 $ & $103.60\pm 0.32 $  \cr
& $\tau$ 		& $0.086 \pm 0.015$ & $0.087 \pm 0.015$ & $0.084 \pm 0.014$ & $0.085 \pm 0.014$ & $0.084 \pm 0.014$  \cr
& $n_s$ 		& $1.002 \pm 0.028$ & $0.965 \pm 0.026$ & $0.967 \pm 0.014$ & $0.980 \pm 0.016$ & $0.967 \pm 0.013$ \cr
&$\ln{(10^{10} \As)}$   & $3.173 \pm 0.060$ & $3.062 \pm 0.047$ & $3.107 \pm 0.036$ & $3.133 \pm 0.042$ & $3.071 \pm 0.030$  \cr
\hline
Extended & $\neff$ 	& $>2.885$	    & $3.192 \pm 1.214$ & $3.623 \pm 0.432$ & $4.608 \pm 0.860$ & $3.454 \pm 0.386$ \cr
& $w$ 			& $-0.794\pm 0.240$ & $-1.076\pm 0.233$ & $-1.054\pm 0.130$ & $-0.986\pm 0.077$ & $-0.937\pm 0.060$ \cr
\hline
Derived & $\Om$ 	& $0.354 \pm 0.078$ & $0.245 \pm 0.044$ & $0.282 \pm 0.019$ & $0.291 \pm 0.016$ & $0.278 \pm 0.013$ \cr 
& $\sigma_8$ 		& $0.849 \pm 0.066$ & $0.807 \pm 0.050$ & $0.865 \pm 0.056$ & $0.876 \pm 0.048$ & $0.785 \pm 0.025$ \cr
& $H_0/100$ 		& $0.742 \pm 0.037$ & $0.740 \pm 0.037$ & $0.723 \pm 0.026$ & $0.748 \pm 0.033$ & $0.703 \pm 0.018$ \cr
\hline
\hline
\end{tabular}
\begin{tablenotes}
\item Same as Table~\ref{table:lam_mnu}, but here $\neff$ and $w$ are instead
set free, with massless neutrinos. As illustrated in Fig.~\ref{fig:w_Neff},
the upper limit of $\neff$ for CMB is severely affected by the preempted prior ($\neff < 10$),
therefore not trustable. Thus we present its 95\% lower limit instead.
\end{tablenotes}
\label{table:w_Neff}
\end{center}
\end{table*}

\begin{table*}\footnotesize
\begin{center}
\caption{Constraints on Parameter Scenario of ``Vanilla+$w$+$\mnu$+$\neff$'' Using Multiple Combinations of Cosmological Probes.}
\begin{tabular}{lcccccccccccccc|}
\hline \hline
&  & CMB & CMB+WL & CMB+BAO+OHD & CMB+BAO+SNIa & CMB+WL+BAO+OHD+SNIa \\
\hline \hline
Vanilla & $100\Obh$	& $2.191 \pm 0.062$ & $2.191 \pm 0.061$ & $2.197 \pm 0.057$ & $2.211 \pm 0.055$ & $2.202 \pm 0.056$ \cr
& $100\Och$		& $16.40 \pm 3.61 $ & $13.35 \pm 2.40 $ & $13.08 \pm 0.93 $ & $14.07 \pm 1.51 $ & $12.70 \pm 0.86 $  \cr
& $10^4\thA$		& $103.24\pm 0.45 $ & $103.51\pm 0.47 $ & $103.53\pm 0.31 $ & $103.37\pm 0.36 $ & $103.52\pm 0.32 $  \cr
& $\tau$ 		& $0.085 \pm 0.015$ & $0.084 \pm 0.015$ & $0.086 \pm 0.014$ & $0.087 \pm 0.014$ & $0.086 \pm 0.014$  \cr
& $n_s$ 		& $0.985 \pm 0.028$ & $0.967 \pm 0.024$ & $0.966 \pm 0.014$ & $0.980 \pm 0.015$ & $0.970 \pm 0.013$ \cr
&$\ln{(10^{10} \As)}$   & $3.146 \pm 0.062$ & $3.088 \pm 0.046$ & $3.104 \pm 0.036$ & $3.126 \pm 0.041$ & $3.089 \pm 0.032$  \cr
\hline
Extended & $\mnu$ [eV]  & $<1.606$ 	    & $<1.539$ 		& $<1.035$ 	    & $<0.869$ 	& $0.556_{-0.288}^{+0.231}$  \cr
	 & $\neff$ 	& $>2.642$	    & $4.125 \pm 1.426$ & $3.794 \pm 0.447$ & $4.574 \pm 0.896$ & $3.839 \pm 0.452$ \cr
	 & $w$ 		& $-1.051\pm 0.335$ & $-1.155\pm 0.278$ & $-1.206\pm 0.183$ & $-1.038\pm 0.088$ & $-1.058\pm 0.088$ \cr
\hline
Derived & $\Om$ 	& $0.338 \pm 0.075$ & $0.292 \pm 0.057$ & $0.282 \pm 0.021$ & $0.295 \pm 0.016$ & $0.294 \pm 0.016$ \cr 
& $\sigma_8$ 		& $0.779 \pm 0.070$ & $0.729 \pm 0.058$ & $0.790 \pm 0.067$ & $0.802 \pm 0.065$ & $0.723 \pm 0.036$ \cr
& $H_0/100$ 		& $0.745 \pm 0.038$ & $0.733 \pm 0.038$ & $0.737 \pm 0.031$ & $0.743 \pm 0.031$ & $0.712 \pm 0.020$ \cr
\hline
\hline
\end{tabular}
\begin{tablenotes}
\item Same as Table~\ref{table:lam_mnu}, yet all members of the extended set ($\mnu,~\neff,~w$) 
are treated as free parameters. Likewise, for CMB, the 95\% lower limit is given for $\neff$. Also the result
of $\mnu$ for the full combination is represented by the best-fit along with 68\% confidence region.  
Note that the case where the entire extended set is kept free against the full combination is 
quoted as ``ALL'' in Section~\ref{sec:discus}. 
\end{tablenotes}
\label{table:w_mnuNeff}
\end{center}
\end{table*}

\end{document}